\documentclass[11pt,reqno,a4paper]{amsart}
\usepackage{amsmath,amssymb,dsfont,graphics,cite,rotating,color}
\usepackage[draft,ulem=normalem]{changes}
\usepackage{mathtools}
\usepackage{microtype}
\def\beq{\begin{equation}}
\def\eeq{\end{equation}}
\def\bea{\begin{eqnarray}}
\def\eea{\end{eqnarray}}

\newtheorem{prop}{Proposition}
\newtheorem{definition}{Definition}

\newtheorem{cor}{Corollary}

\makeatletter
\expandafter\let\expandafter
\reset@font\csname reset@font\endcsname
\def\subeqnarray{\arraycolsep1pt
    \def\@eqnnum\stepcounter##1{\stepcounter{subequation}
        {\reset@font\rm(\theequation\alph{subequation})}}
\jot5mm     \eqnarray}

\makeatother
\newcommand{\CO}{{\mathcal O}}
\newcommand{\CB}{{\mathcal B}}
\newcommand{\CN}{{\mathcal N}}

\newcommand{\CW}{{\mathcal W}}
\newcommand{\CR}{{\mathcal R}}

\newcommand{\CQ}{{\mathcal Q}}

\newcommand{\CC}{{\mathcal C}}
\newcommand{\CA}{{\mathcal A}}
\newcommand{\fref}[1]{\textnormal{(\ref{#1})}}

\def\ep{\varepsilon}

\def\ri{{\rm{i}}}

\def\endpf{\begin{flushright}$\square$\end{flushright}}

\def\su2{{\mathfrak {su}}(2)}
\def\e3{{\mathfrak {e}}(3)}

\begin{document}
\title{High order multiscale analysis of discrete integrable equations}
\author[R. HERN\'ANDEZ HEREDERO, D. LEVI and C. SCIMITERNA]
{R. HERN\'ANDEZ HEREDERO${}^\dag$, \underline{D. LEVI}${}^\diamond$ and C. SCIMITERNA${}^{\ast}$}

\begin{abstract}
In this article we present the results obtained applying the multiple scale expansion up to the order $\ep^6$ to a dispersive multilinear class of equations on a square lattice depending on~13 parameters. We show that the integrability conditions given by the multiple scale expansion give rise to 4 nonlinear equations, 3 of which seem to be new, depending at most on 2 parameters. 
\end{abstract}

\maketitle

{\footnotesize{

\centerline{\it ${}^\dag$
Departamento de Matem\'atica Aplicada a las TIC,} 
\centerline{\it Universidad Polit\'ecnica de Madrid,}
\centerline{\it
ETSI de Sistemas de Telecomunicaci\'on,}
\centerline{\it C. Nikola Tesla s/n, Campus Sur UPM, 28031 Madrid, Spain}
\centerline{e-mail: \texttt{rafael.hernandez.heredero@upm.es}}

\vspace{.3truecm}

\centerline{\it ${}^\diamond$Dipartimento di Matematica e Fisica,}
\centerline{\it Universit\`a degli Studi Roma Tre and Sezione INFN, Roma Tre,}
\centerline{\it Via della Vasca Navale 84, 00146 Roma, Italy}

\vspace{.3truecm}

\centerline{\it ${}^\ast$Istituto di Istruzione Superiore Sansi-Leonardi-Volta,}
\centerline{\it Piazza Carducci, 1 – 06049 Spoleto (PG), Italy}
\centerline{e-mail: \texttt{christian.scimiterna@liceospoleto.edu.it}}

}}

\section{Introduction} \label{intro}
Discrete---or difference---equations play an important role in Mathematical Physics for their double role. First, discrete space-time seems to be basic in the description of fundamental phenomena of nature, as suggested by quantum gravity. On the other hand,  discrete equations are related to differential difference and differential equations through continuous limits. A well-known classification of integrable partial difference equations was given by Adler, Bobenko and Suris~\cite{abs1} in the particular case of equations defined on four lattice points. They used  the ``consistency around the cube'' condition with some symmetry constrains to be able to get definite results. Due to the constraints introduced, this classification is partial and already new equations with respect to those contained in the ABS classification have been found \cite{Viallet,SGR,LY,GY,GH,A}. 

In this paper we provide necessary conditions for the integrability of a class of real, autonomous difference equations in the variable $u: \mathbb Z^2
\rightarrow \mathbb R$  defined on a~$\mathbb{Z}^2$ square-lattice
\beq\label{e}
\CQ (u_{n,m},u_{n+ 1,m},u_{n,m + 1}, u_{n + 1,m + 1}; \beta_1, \beta_2,...)=0,\quad n,m\in\mathbb{Z},
\eeq
where the $\beta_i$'s are real, independent parameters. Integrability  conditions will be determined through a multiscale perturbative development, continuing with the theory explained in references such as~\cite{CD,DMS,DP,Dg,Hls} applicable in differential and difference equations. This approach has the distinctive advantage of providing criteria in a manner completely independent from other current approaches. Multiscale developments can be used to reinforce, enhance or augment our previous knowledge of discrete integrable systems given by other techniques.

We will assume, as in~\cite{abs1}, that (\ref{e}) is linear-affine in every variable, implying that the equation is invariant under the M\"obius transformation $T$
\begin{equation}\label{eqMob}
u_{n,m}\overset{T}{\mapsto} u_{n,m}'=\frac{Au_{n,m}+B}{Cu_{n,m}+D}.
\end{equation}
In this case, (\ref{e}) reduces to a polynomial equation in its variables with an at most fourth order nonlinearity
\bea\label{eq}
\CQ&=&f_0+a_{00}\, u_{00}+ a_{01}\, u_{01} + a_{10}\, u_{10} + a_{11}\, u_{11}+ 
 (\alpha_1{-}\alpha_2) u_{00}\, u_{10} + (\beta_1{-}\beta_2) u_{00}\, u_{01}
 \\ \nonumber
 && \qquad + d_1 u_{00}\, u_{11}+ d_2\, u_{01}\, u_{10} + 
 (\beta_1{+}\beta_2)\, u_{10}\, u_{11} +
  (\alpha_1{+}\alpha_2)\, u_{01}\, u_{11}\\ \nonumber
&& \qquad + 
 (\tau_1{-}\tau_3) u_{00}\, u_{01}\, u_{10}+ (\tau_1{+}\tau_3) u_{00}\, u_{10}\, u_{11} 
+ (\tau_2{+}\tau_4) u_{00}\, u_{01}\, u_{11}
\\ \nonumber
&& \qquad  + 
 (\tau_2{-}\tau_4)\, u_{10}\, u_{01}\, u_{11} 
+ f_1\, u_{00}\, u_{01}\, u_{10}\, u_{11}=0,
\eea
where all coefficients are taken to be real and independent of $n$ and $m$.  
We consider a multiple scale expansion around the dispersive solution 
\bea \label{e2}
u_{n,m} = K^n \Omega^m,
\eea
of the linearized equation of (\ref{eq}). Rewriting the constants~$K$ and~$\Omega$ as $K=e^{\ri k}$ and $\Omega=e^{-\ri\omega}$, and introducing the solution (\ref{e2}) into the linear part of 
Eq.~(\ref{eq}) we get a dispersion relation~$\omega=\omega\left(k\right)$
\bea \label{e3}
\omega=\arctan \left[ \frac{a_{00} a_{01} + a_{10} a_{11} +(a_{00} a_{11} + a_{01} a_{10} ) \cos(k)}{(a_{00} a_{11} - a_{10} a_{01}) \sin(k)} \right], 
\eea
if $f_0=0$. 
The solution (\ref{e2}) of (\ref{eq}) with $f_0=0$ is dispersive if~$\omega(k)$ is a real nonlinear function of the wave number $k$. This leads to the  constraint
\bea \label{e4}
 a_{00}^2 -a_{01}^2+a_{10}^2-a_{11}^2 +2(a_{00} a_{10}-a_{01} a_{11}) \cos(k) = 0.
\eea
The constraint (\ref{e4}) implies that one of the following two conditions must be satisfied: 
\begin{enumerate}
\item $a_{00}=a_{11}\equiv a_1$, $a_{01}=a_{10}\equiv a_2$,
\item $a_{00}=-a_{11}\equiv a_1$, $a_{01}=-a_{10}\equiv a_2$.
\end{enumerate}
Then the dispersion relation (\ref{e3}) reduces to:
\bea \label{e5+}
\omega_{\pm}(k) =\arctan\left[\pm\frac{2a_1 a_2 \pm (a_1^2+a_2^2)\cos(k)}{\left(a_1^2-a_2^2\right)\sin(k)}\right]
\eea
We denote the family of equations~(\ref{eq}) satisfying condition~(1) with dispersion relation $\omega_{+}(k)$ as~$\CQ^+$ and the one with dispersion relation $\omega_{-}(k)$ as~$\CQ^-$. In all the cases $a_1$ and ~$a_2$ cannot be zero and their ratio cannot be equal to $\pm 1$ in order to have a nontrivial dispersion relation.

We will consider integrability conditions for the class of equations $\CQ^+$. The study of the class $\CQ^-$  is left to a future work. 
The result of this work are a series of integrability theorems and a table  of equations, invariant under a restricted  M\"obius transformations, passing the very stringent integrability  conditions obtained with the multiple scale expansion up to $\ep^6$ order.

In Section \ref{s1} we present the main result on the discrete multiscale integrability test,  the conditions up to order $\ep^6$. In Section \ref{s2} we apply it to the classification of dispersive multilinear equations defined on a square lattice $\CQ^+$. Section \ref{s3} is devoted to some conclusive remarks.

\section{The discrete multiscale integrability test}\label{s1}

Consider a dispersive discrete equation of the form $\CQ^+$, i.e.~a completely discrete multilinear dispersive equation defined on a lattice of four points.   In such a situation 
the discrete multiscale integrability test  may be summarized as follows.
\begin{itemize}

\item[\textbf{i.}] One considers a small amplitude solution of Eq.~(\ref{eq}) given by 
 $u_{n,m}=\ep w_{n,m}$, $0 < |\ep| \ll 1$. Then (\ref{eq}) splits into linear and nonlinear terms:
\beq
\CQ^+= \sum_{i=1}^N \ep^i \CQ_i=0, \label{exp}
\eeq
where $N\in\mathbb{N}$ is the nonlinearity order. 
A multilinear equation defined on a square can be at most quartic, i.e.~$N \le 4$. 
In the formal expansion (\ref{exp}) each term $\CQ_i$ contains only
homogeneous polynomials of degree $i$ in~$w_{n,m}$.  If the discrete equation is dispersive then the linear part $\CQ_1$ admits a solution~$w_{n,m}=\exp [\ri( \kappa n -\omega m)]=K^n \Omega^m$,
where
$\omega=\omega(\kappa)=\omega_+(\kappa)$, the dispersion relation, is a real function of $\kappa$ given by Eq.~(\ref{e5+}).\\

\item[\textbf{ii.}] The multiscale expansion of the basic field variable $w_{n,m}$ around the harmonic $K^n \Omega^m$ reads
\beq
w_{n,m}= \sum_{\ell=0}^{\infty} \ep^\ell\sum_{\alpha=-\ell-1}^{\ell+1} 
K^{\alpha n} \Omega^{\alpha m}  u_{\ell+1}^{(\alpha)}, 
\label{bas}
\eeq
where $u^{(\alpha)}_\ell =u^{(\alpha)}_\ell (n_1, \{m_j\})$ is a bounded slowly varying function of its arguments and $u^{(-\alpha)}_{\ell}=\bar u^{(\alpha)}_{\ell}$,
$\bar u_{\ell}$ being the complex conjugate of $u_{\ell}$, because we look only for real solutions.
Here $n_1= \ep n$, $m_j = \ep^j m$ $j= 1, 2, \dots$  are the slow-varying lattice 
variables.\\

\item[\textbf{iii.}] The nearest-neighbors fields are expanded according to the following formulas:
\bea
&& w_{n + 1,m} =  \sum_{\ell=0}^{\infty} \ep^\ell \sum_{\alpha=-\ell-1}^{\ell+1}
K^{\alpha (n+ 1)} \Omega^{\alpha m}
\sum_{j= {\rm max} (0, |\alpha|-1)}^{\ell}
 \CA_{ \ell- j}  u_{j+1}^{(\alpha)} , \label{n} \\
&& w_{n ,m + 1} =  \sum_{\ell=0}^{\infty} \ep^\ell \sum_{\alpha=-\ell-1}^{\ell+1}
 K^{\alpha n} \Omega^{\alpha (m - 1)}
\sum_{j= {\rm max} (0, |\alpha|-1)}^{\ell}
\CB_{ \ell- j}  u_{j+1}^{(\alpha)} , \label{m} \\
&& w_{n + 1 ,m + 1} =  \sum_{\ell=0}^{\infty} \ep^\ell \sum_{\alpha=-\ell-1}^{\ell+1}
 K^{\alpha (n + 1)} \Omega^{  \alpha (m - 1)}
\sum_{j= {\rm max} (0, |\alpha|-1)}^{\ell}
\CC_{ \ell- j}  u_{j+1}^{(\alpha)} . \label{nm} 
\eea
The operators $\CA_i$, $\CB_i$, $\CC_i$, 
are equal to 1 when $i=0$, and for some lower values of~$i$ are:\\
$$\begin{array}{||c||c|c|c|c||}\hline 
& i=1 & i=2 & i=3& i=4\\ \hline\hline 
& & & & \\ 
\CA_i  &  \delta_{n_1} & \frac12\delta_{n_1}^2 & \frac16\delta_{n_1}^3 & \frac{1}{24}\delta_{n_1}^4\\
& & & & \\ \hline
& & & & \\ 
\CB_i  &  \delta_{m_1} &\frac12\delta_{m_1}^2 + \delta_{m_2} & 
\frac16\delta_{m_1}^3+\delta_{m_1}\delta_{m_2}+\delta_{m_3} & \frac{1}{24}\delta_{m_1}^4+\frac12\delta_{m_1}^2\delta_{m_2}+\frac12\delta_{m_2}^2+\delta_{m_1}\delta_{m_3}+\delta_{m_4}\\
& & & & \\ \hline 
& & & & \\ 
\CC_i  & \nabla & \frac12\nabla^2+\delta_{m_2} &
\frac16 \nabla^3+\nabla\delta_{m_2}+\delta_{m_3} & \frac{1}{24}\nabla^4+\frac12\nabla^2\delta_{m_2}+\frac12\delta_{m_2}^2+\nabla\delta_{m_3}+\delta_{m_4}\\
& & & & \\ \hline 
\end{array}
$$\\
where  $\delta_{k}$ are the formal derivatives with respect to the index $k$, $\delta_k\coloneqq\partial_k$ and $\nabla\coloneqq\delta_{m_1}+\delta_{n_1}$. The operator $\delta_k$ can always be expressed in terms of powers of the difference operators by the well known identity
$$
\delta_{k}= \sum_{i=1}^\infty
\frac{(-1)^{i-1}}{i}\Delta_{k}^i, 
$$
where $\Delta_{k}$ is the discrete first right difference
operator with respect to the variable $k$, i.e.
$\Delta_k u_k \coloneqq u_{k+1} - u_k$.

A function $f_k$  is a {\it slow-varying} function of order $L$  if $\Delta_k^{L+1} f_k=0$.  The $\delta_k$-operators, which in principle are formal infinite series in powers of $\Delta_k$, when acting on {\it slow-varying} functions of finite
order $L$ reduce to polynomials in $\Delta_k$ at most of order $L$. 
We shall assume that we are dealing with functions of an infinite slow-varying order, i.e.~$L=\infty$,
so the $\delta_k$-operators 
may be taken as differential operators acting on the indices of the harmonics $u_j^{(\alpha)}$.\\

\item[\textbf{iv.}] Substituting the expansions (\ref{bas}-\ref{nm}) into (\ref{exp}), we get an equation of the following
form:
\beq
\sum_j \ep^j \sum_{\alpha} \CW_j^{(\alpha)}K^{\alpha n} \Omega^{\alpha m} =0, \label{eee}
\eeq
i.e.~we must have $\CW_j^{(\alpha)}=0$ for all $\alpha$ and $j$. 
Notice that the equations $\CW_j^{(\alpha)}=0$ are equations for the slowly varying functions $u_{\ell+1}^{(\alpha)}$ with $\ell \leq j$.
\end{itemize}

The multiscale expansion of the $\CQ^+$ equation for functions of infinite order thus gives rise to a \emph{system of continuous partial differential equations.} At the lowest order (slow-time $m_{2}$) one gets a Nonlinear Schr\"odinger equation (\emph{NLS}) for the first harmonic $u^{\left(1\right)}_{1}$. We will use orders beyond that to define the values of the constants appearing in $\CQ^+$ for which the equation is integrable. The first attempt to go beyond the \emph{NLS} order in the case of partial differential equations was presented by Santini, Degasperis and Manakov in \cite{DMS} and by Kodama and Mikhailov using normal forms\cite{KM}. In \cite{DMS} the authors, starting from $S$-integrable models, through a combination of an asymptotic functional analysis and spectral methods, succeeded in removing all the secular terms from the reduced equations, order by order. Their results could be summarized in the following statements:

\begin{enumerate}
\item The number of slow-time variables required for the amplitudes $u^{\left(\alpha\right)}_{j}$ coincides with the number of non-vanishing coefficients $\omega_{j}\left(k\right)=\frac{1}{j!}\frac{d^j \omega(k)}{dk^j}$;

\item The amplitude $u^{\left(1\right)}_{1}$ evolves at the slow-times $t_{\sigma}\coloneqq m_{\sigma}$, $\sigma\geq 3$ according to the $\sigma-$th equation of the \emph{NLS} hierarchy;

\item The amplitudes of the higher perturbations of the first harmonic $u^{\left(1\right)}_{j}$, $j\geq 2$ evolve at the slow-times $t_{\sigma}$, $\sigma\geq 2$ according to certain {linear, nonhomogeneous} equations when taking into account some {asymptotic boundary conditions}.
\end{enumerate}
From these statements one can conclude that the cancellation at each stage of the perturbation process of all the secular terms  is a {\it sufficient condition} to uniquely fix the evolution equations followed by every $u^{\left(1\right)}_{j}$, $j\geq 1$ for each slow-time $t_\sigma$. Conversely, the results in~\cite{DP} imply that this expansion is secularity-free. Thus, this procedure provides \emph{necessary and sufficient} conditions to get secularity-free reduced equations. Following \cite{DP} we can state the following proposition:
\begin{prop}
If a nonlinear dispersive partial difference equation  is integrable, then under a multiscale expansion the functions $u^{\left(1\right)}_{l}$, $l\geq1$ satisfy the equations
\begin{subequations}\label{Valentia}
\begin{gather}
\partial_{t_{\sigma}}u^{\left(1\right)}_{1}=K_{\sigma}\left[u^{\left(1\right)}_{1}\right],\label{Valentia1}\\
M_{\sigma}u^{\left(1\right)}_{j}=f_{\sigma}(j),\quad M_{\sigma}\coloneqq\partial_{t_{\sigma}}-K_{\sigma}^{\prime}\left[u^{\left(1\right)}_{1}\right],\label{Valentia2}
\end{gather}
\end{subequations}
 $\forall\ j,\ \sigma\geq 2$, where $K_{\sigma}\left[u^{\left(1\right)}_{1}\right]$ is the $\sigma$--th flow in the nonlinear Schr\" odinger hierarchy. All the other  $u_{j}^{(\kappa)}$, $\kappa\geq 2$ are expressed in terms of differential monomials of $u_{\rho}^{(1)}$, $\rho\leq j$.
 \end{prop}
In  (\ref{Valentia2})   $f_{\sigma}(j)$ is a nonhomogeneous \emph{nonlinear} forcing term depending on all the $u^{(1)}_{\kappa}$, $1\leq\kappa\leq j-1$, their complex conjugates and their $\xi$-derivatives, where $\xi$ is a variable representing the group velocity and expressed through a linear combination of the slow-space and the first slow-time $t_{1}$, while $K_{\sigma}^{\prime}\left[u\right]v$ is the Frechet derivative of the nonlinear term $K_{\sigma}[u]$ along the direction $v$ defined by $K_{\sigma}^{\prime}[u]v\coloneqq\frac{d} {ds}K_{\sigma}[u+sv]\mid_{s=0}$, 
i.e.~the linearization  of the expression $K_{\sigma}[u]$ along the direction $v$ near the function $u$.

In order to characterize the flows~$K_{\sigma}\left[u^{\left(1\right)}_{1}\right]$ and the nonlinear forcing terms $f_{\sigma}(j)$, following \cite{Dg}, we introduce the finite dimensional vector spaces $\mathcal{P}_{\ell}$, $\ell\geq 2$, as being the set of all homogeneous, fully-nonlinear, differential polynomials in the functions $u_{j}^{(1)}$, $j\geq 1$, their complex conjugates and their $\xi$-derivatives of homogeneity order $\ell$  in $\ep$ and $1$ in the accompanying exponential~$e^{\ri\theta}=e^{\ri( \kappa n -\omega m)}$, where
\[
\mbox{order}_{\ep}\left(\partial_{\xi}^{\kappa}u^{(1)}_{j}\right)=\mbox{order}_{\ep}\left(\partial_{\xi}^{\kappa}\bar u^{(1)}_{j}\right)=\kappa+j,\quad \kappa\geq 0.
\] 
We introduce the subspaces $\mathcal{P}_{ \ell}(\jmath)$ of $\mathcal{P}_{\ell}$, $\jmath\geq 1$, $\ell\geq 2$, whose elements are homogeneous, fully-nonlinear, differential polynomials in the functions $u_{k}^{(1)}$, their complex conjugates and their $\xi$-derivatives with $1\leq k\leq \jmath$. Firstly from these definitions it follows that $\mathcal{P}_{\ell}=\mathcal{P}_{\ell}\left(\ell-2\right)$, that is $\jmath\leq \ell-2$. In fact the terms $u^{(1)}_{\ell}$ and $\bar u^{(1)}_{\ell}$, as well as $\partial_{\xi}u^{(1)}_{\ell-1}$ and $\partial_{\xi}\bar u^{(1)}_{\ell-1}$, are not included in~$\mathcal{P}_{\ell}$ as any monomial should enter nonlinearly and terms like $u^{(1)}_{\ell-1}$ and $\bar u^{(1)}_{\ell-1}$  cannot be combined with any other of the monomials $u^{(1)}_{1}$ or $\bar u^{(1)}_{1}$ to give the right homogeneity degree in $e^{\ri\theta}$. For the same reasons, terms of the types $\partial_{\xi}^{\kappa}u^{(1)}_{\ell-\kappa}$, $\partial_{\xi}^{\kappa}\bar u^{(1)}_{\ell-\kappa}$, $0\leq\kappa\leq \ell-1$ and $\partial_{\xi}^{\kappa}u^{(1)}_{\ell-\kappa-1}$, $\partial_{\xi}^{\kappa}\bar u^{(1)}_{\ell-\kappa-1}$, $0\leq\kappa\leq \ell-2$ cannot appear. So the space $\mathcal{P}_{\ell}(\jmath)$  is defined as that functional space generated by the base of monomials of the following types
\begin{eqnarray}
\prod_{\alpha,\beta,\gamma,\delta}\left(\partial_{\xi}^{\alpha}u^{(1)}_{\beta}\right)^{\rho\left(\alpha,\beta\right)}\left(\partial_{\xi}^{\gamma}\bar u^{(1)}_{\delta}\right)^{\sigma\left(\gamma,\delta\right)},\ \ \ \rho\left(\alpha,\beta\right)\geq 0,\ \ \forall\alpha,\beta,\ \ \ \sigma\left(\gamma,\delta\right)\geq 0,\ \ \forall\gamma,\delta,\nonumber
\end{eqnarray}
where the product is extended for $1\leq\beta,\delta\leq \jmath\leq \ell-2$, $0\leq\alpha\leq \ell-\beta-2$ and $0\leq\gamma\leq \ell-\delta-2$, so that
\[
\sum_{\alpha,\beta,\gamma,\delta}\left(\alpha+\beta\right)\rho\left(\alpha,\beta\right)+\left(\gamma+\delta\right)\sigma\left(\gamma,\delta\right)=\ell,\quad \sum_{\alpha,\beta,\gamma,\delta}\rho\left(\alpha,\beta\right)-\sigma\left(\gamma,\delta\right)=1\nonumber.
\]
For $n\geq 3$ the subspaces $\mathcal{P}_{\ell}(\jmath)$, can be generated recursively starting from the lowest one, corresponding to $\ell=2$ by the following relation
\begin{eqnarray}
\mathcal{P}_{\ell}(\jmath)=\partial_{\xi}\mathcal{P}_{\ell-1}(\jmath)\cup\left\{\prod_{\beta,\delta}\left(u^{(1)}_{\beta}\right)^{\rho\left(\beta\right)}\left(\bar u^{(1)}_{\delta}\right)^{\sigma\left(\delta\right)}\right\},\nonumber 
\end{eqnarray}
where $\rho\left(\beta\right)\geq 0$ $\forall\beta$, $\sigma\left(\delta\right)\geq 0$ $\forall\delta$ and the product is extended for $1\leq\beta,\delta\leq \jmath\leq \ell-2$, so that
\begin{eqnarray}
\sum_{\beta,\delta}\beta\rho\left(\beta\right)+\delta\sigma\left(\delta\right)=\ell,\ \ \ \sum_{\beta,\delta}\rho\left(\beta\right)-\sigma\left(\delta\right)=1.\nonumber
\end{eqnarray}
It is then clear that in general $K_{n}\left[u^{\left(1\right)}_{1}\right]\in\left\{\partial_{\xi}^{\ell}u^{\left(1\right)}_{1}\right\}\cup\mathcal{P}_{\ell+1}(1)$ and that $f_{\sigma}(j)\in\mathcal{P}_{\sigma+j}(j-1)$, $\forall\sigma$, $j\geq 2$.  

Eqs.~(\ref{Valentia}) are a \emph{necessary} condition for integrability and represent a hierarchy of \emph{compatible} evolutions for the same function $u^{\left(1\right)}_{1}$ at different slow-times.  The compatibility of Eqs.~(\ref{Valentia2})  implies some commutativity conditions among their r.h.s.~terms $f_{\sigma}(j)$. If they are satisfied the operators $M_{\sigma}$ defined in Eq.~(\ref{Valentia2}) commute among themselves. 
Once we fix the index $j\geq 2$ in the set of Eqs.~(\ref{Valentia2}), this commutativity condition implies the following \emph{compatibility} conditions
\begin{eqnarray}
M_{\sigma}f_{\sigma'}\left(j\right)=M_{\sigma'}f_{\sigma}\left(j\right),\ \ \ \forall\, \sigma,\sigma'\geq 2,\label{Lavinia}
\end{eqnarray}
where, as $f_{\sigma}\left(j\right)$ and $f_{\sigma'}\left(j\right)$ are functions of the different perturbations of the fundamental harmonic up to degree $j-1$, the time derivatives $\partial_{t_{\sigma}}$, $\partial_{t_{\sigma'}}$ of those harmonics appearing respectively in $M_{\sigma}$ and $M_{\sigma'}$ have to be eliminated using the evolution equations (\ref{Valentia}) up to the index $j-1$. The commutativity conditions (\ref{Lavinia}) turn out to be an {\bf integrability test}.

 We finally define the {\bf degree of integrability} of a given equation:
\begin{definition}\label{FrancescoColonnaRomano}
{\it If the relations (\ref{Lavinia}) are satisfied up to the index $j$, $j\geq 2$, we say that our equation is asymptotically integrable of degree $j$ or $A_{j}$ integrable.}
\end{definition}

Conjecturing that an $A_{\infty}$ degree of asymptotic integrability actually implies integrability, we have that under this assumption the relations (\ref{Valentia}, \ref{Lavinia}) are a \emph{sufficient} condition for the $S$-integrability or that integrability is a \emph{necessary} condition to have a multiscale expansion where all the Eqs.~(\ref{Valentia}) are satisfied. So the multiscale integrability test tell us that $\CQ^+$ will be integrable if its multiscale expansion will follow all the infinite relations (\ref{Valentia}, \ref{Lavinia}). The higher the degree of asymptotic integrability, the nearer the equation will be to an integrable one. However, as we can test the conditions (\ref{Valentia}, \ref{Lavinia}) only up to a finite order (currently $A_{4}$), from them we can only derive necessary conditions for integrability, so we will not be able to state with certainty that the discrete equation is integrable. The results obtained at a finite but sufficiently high order will have a good probability to correspond to an integrable equation, but we need to use other techniques to prove it with certainty.

Let us present for completeness the lowest order conditions for asymptotic $S$-integrability of order $k$ or $A_{k}$-integrability conditions. To simplify the notation, we will use for $u^{\left(1\right)}_{j}$ the concise form $u(j)$, $j\geq 1$. Moreover, for the convenience of the reader, we list the fluxes $K_{\sigma}\left[u\right]$ of the \emph{NLS} hierarchy for $u$ up to $\sigma=5$:
\begin{subequations}\label{Rutuli}\allowdisplaybreaks
\begin{gather}
K_{1}[u]\coloneqq Au_{\xi},\\
K_{2}[u]\coloneqq-\ri\rho_{1}\left[u_{\xi\xi}+\frac{\rho_{2}} {\rho_{1}}|u|^2u \right],\label{Rutuli1}\\
K_{3}[u]\coloneqq B\left[u_{\xi\xi\xi}+\frac{3\rho_{2}} {\rho_{1}}|u|^2u_{\xi}\right],\label{Rutuli2}\\
K_{4}[u]\coloneqq-\ri C\left\{u_{\xi\xi\xi\xi}+\frac{\rho_{2}} {\rho_{1}}\left[\frac{3\rho_{2}} {2\rho_{1}}|u|^4u+4|u|^2u_{\xi\xi}+3u_{\xi}^2\bar u+2|u_{\xi}|^2u+u^2\bar u_{\xi\xi}\right]\right\},\label{Rutuli3}\\
K_{5}[u]\coloneqq D\left\{u_{\xi\xi\xi\xi\xi}+\frac{5\rho_{2}}{\rho_{1}}\left[\frac{3\rho_{2}}{2\rho_{1}}|u|^4u_{\xi}+|u_{\xi}|^2u_{\xi}+\left(u\bar u_{\xi}{+}2\bar uu_{\xi}\right)u_{\xi\xi}+uu_{\xi}\bar u_{\xi\xi}+|u|^2u_{\xi\xi\xi}\right]\right\},\label{Rutuli4}
\end{gather}
\end{subequations}
and the corresponding $K_{\sigma}^{'}[u]v$ up to $\sigma=4$:
\begin{subequations}\allowdisplaybreaks
\begin{gather}
K_{1}^{\prime}[u]v= Av_{\xi},\label{Arenta}\\
K_{2}^{\prime}[u]v=-\ri\rho_{1}\left\{v_{\xi\xi}+\frac{\rho_{2}} {\rho_{1}}\left[u^2\bar v+2|u|^2v\right]\right\},\label{ArtemideEfesina}
\\
K_{3}^{\prime}[u]v=B\left\{v_{\xi\xi\xi}+\frac{3\rho_{2}} {\rho_{1}}\left[|u|^2v_{\xi}+\bar uu_{\xi}v+uu_{\xi}\bar v\right]\right\},\label{Abruzzo3}
\\
K_{4}^{\prime}\left[u\right]v=-iC\left\{v_{\xi\xi\xi\xi}+\frac{\rho_{2}}{\rho_{1}}\Bigr[u^{2}\bar v_{\xi\xi}+4|u|^{2}v_{\xi\xi}+2uu_{\xi}\bar v_{\xi}+2u\bar u_{\xi}v_{\xi}+6\bar u u_{\xi}v_{\xi}+\right.\\
\nonumber\qquad \left.{}+4uu_{\xi\xi}\bar v+3u_{\xi}^{2}\bar v+\frac{3\rho_{2}}{\rho_{1}}|u|^{2}u^{2}\bar v+4\bar u u_{\xi\xi}v+2u\bar u_{\xi\xi}v+\frac{9\rho_{2}}{2\rho_{1}}|u|^{4}v+2|u_{\xi}|^{2}v\Bigr]\right\},
\end{gather}
\end{subequations}
where $A$, $\rho_{1}$, $\rho_{2}$, $B$, $C$ and $D$ are all non null and, if $\rho_{2}\not=0$, real arbitrary constants.

\subsection{The $A_{1}$-integrability condition.}\ \\

The $A_{1}$-integrability condition is given by the reality of the coefficient $\rho_{2}$ of the nonlinear term in the \emph{NLS}. It is obtained commuting the \emph{NLS} flux $K_{2}[u]$ with the flux $B\left[u_{\xi\xi\xi}+\tau |u|^2u_{\xi}+\mu u^2\bar u_{\xi}\right]$ with $\tau$ and $\mu$ constants. This commutativity condition gives, if $\rho_{2}\not =0$, 
\begin{eqnarray}
\textrm{Im}\left[\rho_{2}\right]=\textrm{Im}\left[B\right]=\textrm{Im}\left[\rho_{1}\right]=0,\ \ \tau=3\rho_{2}/\rho_{1},\ \ \mu=0.\label{Montesiepi}
\end{eqnarray}

We remark that, when $\rho_{2}\not=0$, by the same method it is possible to determine all the coefficients of all the higher \emph{NLS}-symmetries (\ref{Rutuli}) together with the reality conditions of the coefficients $A$, $C$ and $D$. 

\subsection{The $A_{2}$-integrability conditions.}\ \\

The $A_{2}$-integrability conditions  are obtained choosing $j=2$ in the compatibility conditions~(\ref{Lavinia}) with $\sigma=2$ and $\sigma'=3$ or alternatively $\sigma'=4$, respectively 
\begin{subequations}
\begin{gather}
M_{2}f_{3}\left(2\right)=M_{3}f_{2}\left(2\right),\label{Turno}\\
M_{2}f_{4}\left(2\right)=M_{4}f_{2}\left(2\right).\label{Turno1}
\end{gather}
\end{subequations}
In this case  $f_{2}(2)$, $f_{3}(2)$ and $f_{4}(2)$ will be identified by respectively two, ($a, b$), five, ($\alpha, \beta, \gamma, \delta, \epsilon$), and eight, ($\theta_1, \cdots, \theta_8$), complex constants
\begin{subequations}
\begin{align}
f_{2}(2)&\coloneqq au_{\xi}(1)|u(1)|^2+b\bar u_{\xi}(1)u(1)^2,\label{Abruzzo1}\\
f_{3}(2)&\coloneqq\alpha |u(1)|^4u(1)+\beta |u_{\xi}(1)|^2u(1)+\gamma u_{\xi}(1)^2\bar u(1)
+\delta\bar u_{\xi\xi}(1)u(1)^2+\epsilon |u(1)|^2u_{\xi\xi}(1),
\label{Abruzzo2}
\\
f_{4}\left(2\right)&\coloneqq\theta_{1}|u\left(1\right)|^{4}u_{\xi}\left(1\right)+\theta_{2}|u\left(1\right)|^{2}u\left(1\right)^{2}\bar u_{\xi}\left(1\right)+\theta_{3}|u_{\xi}\left(1\right)|^{2}u_{\xi}\left(1\right)
\\
&\quad{}+\theta_{4}u\left(1\right)\bar u_{\xi}\left(1\right)u_{\xi\xi}\left(1\right)+\theta_{5}\bar u\left(1\right)u_{\xi}\left(1\right)u_{\xi\xi}\left(1\right)+\theta_{6}u\left(1\right)u_{\xi}\left(1\right)\bar u_{\xi\xi}\left(1\right)\nonumber
\\
&\quad{}+\theta_{7}|u\left(1\right)|^{2}u_{\xi\xi\xi}\left(1\right)+\theta_{8}u\left(1\right)^{2}\bar u_{\xi\xi\xi}\left(1\right).\nonumber
\end{align}
\end{subequations}
As $\rho_{2}\not=0$, eliminating from Eq.~(\ref{Turno}) the derivatives of $u(1)$ with respect to the slow-times $t_{2}$ and $t_{3}$, using the evolutions (\ref{Valentia1}) with $\sigma=2$ and $\sigma'=3$ and equating term by term, we obtain the following two $A_{2}$-integrability conditions
\begin{eqnarray}
a=\bar a,\ \ \ b=\bar b.\label{CieloUrbico}
\end{eqnarray}
 So we have  two  conditions obtained requiring the reality of the coefficients $a$ and $b$. The expressions of $\alpha$, $\beta$, $\alpha$, $\delta$ in terms of $a$ and $b$ are:
\begin{eqnarray}
\alpha=\frac{3\ri Ba\rho_{2}} {4\rho_{1}^2},\ \ \ \beta=\frac{3\ri Bb} {\rho_{1}},\ \ \ \gamma=\frac{3\ri Ba} {2\rho_{1}},\ \ \ \delta=0,\ \ \ \epsilon=\gamma.\label{Molise}
\end{eqnarray}
The same integrability conditions (\ref{CieloUrbico}) can be derived using Eq.~(\ref{Turno1}). As in our analysis we will need them, here follows the explicit expressions of the coefficients of the forcing term $f_{4}\left(2\right)$
\begin{equation}
\begin{gathered}
\theta_{1}=\frac{6Ca\rho_{2}}{\rho_{1}^2},\ \ \ \theta_{2}=\frac{3Cb\rho_{2}}{\rho_{1}^2},\ \ \ \theta_{3}=\frac{\left(a+3b\right)C}{\rho_{1}},\ \ \ \theta_{4}=\frac{\left(a+4b\right)C}{\rho_{1}},
\\
\theta_{5}=\frac{5Ca}{\rho_{1}},\ \ \ \theta_{6}=\frac{\left(a+2b\right)C}{\rho_{1}},\ \ \ \theta_{7}=\frac{2Ca}{\rho_{1}},\ \ \ \theta_{8}=\frac{Cb}{\rho_{1}}.
\end{gathered}\label{Moly}
\end{equation}

\subsection{The $A_{3}$-integrability conditions.}\ \\

The $A_{3}$-integrability conditions are derived in a similar way setting $j=3$ in the compatibility conditions (\ref{Lavinia}) with $\sigma=2$ and $\sigma'=3$, so that $M_{2}f_{3}\left(3\right)=M_{3}f_{2}\left(3\right)$. In this case  $f_{2}(3)$ and $f_{3}(3)$ will be respectively identified by 12 and 26 complex constants
\begin{subequations} \label{f23}
\begin{align} 
 f_{2}(3)&\coloneqq\tau_{1}|u(1)|^4u(1)+\tau_{2}|u_{\xi}(1)|^2u(1)+\tau_{3}|u(1)|^2u_{\xi\xi}(1)+\tau_{4}\bar u_{\xi\xi}(1)u(1)^2\label{Lazio4}
\\
&\quad{}+\tau_{7}\bar u_{\xi}(2)u(1)^2+\tau_{8}u(2)^2\bar u(1)+\tau_{9}|u(2)|^2u(1)+\tau_{10}u(2)u_{\xi}(1)\bar u(1)
\nonumber
\\
&\quad{}+\tau_{11}u(2)\bar u_{\xi}(1)u(1)+\tau_{12}\bar u(2)u_{\xi}(1)u(1)+\tau_{5}u_{\xi}(1)^2\bar u(1)+\tau_{6}u_{\xi}(2)|u(1)|^2,\nonumber\\
f_{3}(3)&\coloneqq\gamma_{1}|u(1)|^4u_{\xi}(1)+\gamma_{2}|u(1)|^2u(1)^2\bar u_{\xi}(1)+\gamma_{3}|u(1)|^2u_{\xi\xi\xi}(1)\label{Lazio5}\\
&\quad{}+\gamma_{5}|u_{\xi}(1)|^2u_{\xi}(1)+\gamma_{6}\bar u_{\xi\xi}(1)u_{\xi}(1)u(1)+\gamma_{7}u_{\xi\xi}(1)\bar u_{\xi}(1)u(1)\nonumber\\
&\quad{}+\gamma_{9}|u(1)|^4u(2)+\gamma_{10}|u(1)|^2u(1)^2\bar u(2)+\gamma_{11}\bar u_{\xi}(1)u(2)^2+\gamma_{12}u_{\xi}(1)|u(2)|^2\nonumber\\
&\quad{}+\gamma_{13}|u_{\xi}(1)|^2u(2)+\gamma_{14}|u(2)|^2u(2)+\gamma_{15}u_{\xi}(1)^{2}\bar u(2)+\gamma_{16}|u(1)|^2u_{\xi\xi}(2)\nonumber\\
&\quad{}+\gamma_{17}u(1)^2\bar u_{\xi\xi}(2)+\gamma_{18}u(2)\bar u_{\xi\xi}(1)u(1)+\gamma_{19}u(2)u_{\xi\xi}(1)\bar u(1)\nonumber\\
&\quad{}+\gamma_{21}u(2)u_{\xi}(2)\bar u(1)+\gamma_{22}\bar u(2)u_{\xi}(2)u(1)+\gamma_{23}u_{\xi}(2)u_{\xi}(1)\bar u(1)\nonumber\\
&\quad{}+\gamma_{25}\bar u_{\xi}(2)u_{\xi}(1)u(1)+\gamma_{26}\bar u_{\xi}(2)u(2)u(1)+\gamma_{4}u(1)^2\bar u_{\xi\xi\xi}(1)  \nonumber \\ 
&\quad{}+\gamma_{8}u_{\xi\xi}(1)u_{\xi}(1)\bar u(1)+\gamma_{20}\bar u(2)u_{\xi\xi}(1)u(1)+\gamma_{24}u_{\xi}(2)\bar u_{\xi}(1)u(1).\nonumber
\end{align}
\end{subequations}
Eliminate from Eq.~(\ref{Turno}) with $j=3$ the derivatives of $u(1)$ with respect to the slow-times $t_{2}$ and $t_{3}$ using the evolutions (\ref{Valentia1}) respectively with $\sigma=2$ and $\sigma'=3$ and the derivatives of  $u(2)$ using the evolutions (\ref{Valentia2}) with $\sigma=2$ and $\sigma'=3$. Equating the remaining terms term by term, with~$\rho_{2}\not=0$  and, indicating with $R_{i}$ and $I_{i}$ the real and imaginary parts of $\tau_{i}$, $i=1,\ldots,12$, we obtain the following 15 $A_{3}$-integrability conditions
\begin{gather}\allowdisplaybreaks
R_{1}=-\frac{aI_{6}} {4\rho_{1}},\quad R_{3}=\frac{(b-a)I_{6}} {2\rho_{2}}-\frac{aI_{12}} {2\rho_{2}},\quad R_{4}=\frac{R_{2}} {2}+\frac{(a-b)I_{6}} {4\rho_{2}}+\frac{aI_{12}} {4\rho_{2}},
\nonumber\\
R_{5}=\frac{R_{2}} {2}+\frac{(a-b)I_{6}} {4\rho_{2}}+\frac{(2b-a)I_{12}} {4\rho_{2}},\quad R_{6}=-\frac{aI_{8}} {\rho_{2}},\quad R_{7}=R_{12}+\frac{(a-b)I_{8}} {\rho_{2}},
\nonumber\\
R_{8}=R_{9}=0,\quad R_{10}=R_{12},\quad R_{11}=R_{12}+\frac{(a-2b)I_{8}} {\rho_{2}},
\label{Siculi}\\
I_{4}=\frac{(b+a)R_{12}} {4\rho_{2}}+\frac{\rho_{1}I_{1}} {\rho_{2}}+\frac{I_{2}-I_{3}-2I_{5}} {4}+\frac{\left[2b(a-b)+a^2\right]I_{8}} {4\rho_{2}^2},\quad I_{7}=0,
\nonumber\\
I_{9}=2I_{8},\quad I_{10}=I_{12},\quad I_{11}=I_{6}+I_{12}.
\nonumber
\end{gather}
The expressions of the $\gamma_{j}$, $j=1,\ldots,26$ as functions of the $\tau_{i}$, $i=1,\ldots,12$ are:
\begin{gather}\allowdisplaybreaks
\gamma_{1}=\frac{3B} {8\rho_{1}^2}\left[-2bR_{12}-8\rho_{1}I_{1}+2(I_{2}-2I_{3}-2I_{5})\rho_{2}+\ri (b-5a)I_{6}+\frac{2a^2I_{8}} {\rho_{2}}-3\ri aI_{12}\right],
\nonumber\\
\gamma_{2}=-\frac{3Ba} {4\rho_{1}^2}\left[\ri I_{6}+\frac{(a-2b)I_{8}} {\rho_{2}}+\tau_{12}\right],\ \gamma_{3}=\frac{3\ri B\tau_{3}} {2\rho_{1}},\ \gamma_{4}=0,\ \gamma_{5}=\frac{3\ri B\tau_{2}} {2\rho_{1}},\ \gamma_{6}=\frac{3\ri B\tau_{4}} {\rho_{1}},
\nonumber\\
\gamma_{7}=\gamma_{5},\quad\gamma_{8}=\gamma_{3}+\frac{3\ri B\tau_{5}} {\rho_{1}},\quad\gamma_{9}=-\frac{3B(\rho_{2}I_{6}+3a\ri I_{8})} {4\rho_{1}^2},\quad\gamma_{10}=\frac{3\ri B\rho_{2}R_{6}} {2\rho_{1}^2},\quad \gamma_{11}=0,
\nonumber\\
\gamma_{12}=\frac{3\ri B\tau_{9}} {2\rho_{1}},\quad\gamma_{13}=\frac{3\ri B\tau_{11}} {2\rho_{1}},\quad\gamma_{14}=0,\quad\gamma_{15}=\frac{3\ri B\tau_{12}} {2\rho_{1}},
\quad\gamma_{16}=\frac{3\ri B\tau_{6}} {2\rho_{1}},
\label{gam}\\
\quad\gamma_{17}=\gamma_{18}=0,\quad 
\gamma_{19}=\frac{3\ri B\tau_{10}} {2\rho_{1}},\quad\gamma_{20}=\gamma_{15},\quad\gamma_{21}=\frac{3\ri B\tau_{8}} {\rho_{1}},\quad\gamma_{22}=\gamma_{12},
\nonumber\\
\gamma_{23}=\gamma_{16}+\gamma_{19},\quad\gamma_{24}=\gamma_{13},\quad\gamma_{25}=\frac{3\ri B\tau_{7}} {\rho_{1}},\quad\gamma_{26}=0.
\nonumber
\end{gather}

\subsection{The $A_{4}$-integrability conditions.}\ \\

The $A_{4}$-integrability conditions are derived similarly from (\ref{Lavinia}) with $j=4$, that is $M_{2}f_{3}\left(4\right)=M_{3}f_{2}\left(4\right)$. Now $f_{2}(4)$ and $f_{3}(4)$ are respectively defined by 34 and 77 complex constants
\begin{subequations}\allowdisplaybreaks
\begin{align}
f_{2}\left(4\right)&\coloneqq\eta_{1}|u(1)|^4u_{\xi}(1)+\eta_{2}|u(1)|^2u(1)^2\bar u_{\xi}(1)+\eta_{3}|u(1)|^2u_{\xi\xi\xi}(1)
\label{f24}\\
&\quad+{}\eta_{5}|u_{\xi}(1)|^2u_{\xi}(1)+\eta_{6}\bar u_{\xi\xi}(1)u_{\xi}(1)u(1)+\eta_{7}u_{\xi\xi}(1)\bar u_{\xi}(1)u(1)
\nonumber\\
&\quad+{}\eta_{9}|u(1)|^4u(2)+\eta_{10}|u(1)|^2u(1)^2\bar u(2)+\eta_{11}\bar u_{\xi}(1)u(2)^2+\eta_{12}u_{\xi}(1)|u(2)|^2
\nonumber\\
&\quad+{}\eta_{13}|u_{\xi}(1)|^2u(2)+\eta_{14}|u(2)|^2u(2)+\eta_{15}u_{\xi}(1)^{2}\bar u(2)+\eta_{16}|u(1)|^2u_{\xi\xi}(2)
\nonumber\\
&\quad+{}\eta_{17}u(1)^2\bar u_{\xi\xi}(2)+\eta_{18}u(2)\bar u_{\xi\xi}(1)u(1)+\eta_{19}u(2)u_{\xi\xi}(1)\bar u(1)
\nonumber\\
&\quad+{}\eta_{21}u(2)u_{\xi}(2)\bar u(1)+\eta_{22}\bar u(2)u_{\xi}(2)u(1)+\eta_{23}u_{\xi}(2)u_{\xi}(1)\bar u(1)
\nonumber\\
&\quad+{}\eta_{25}\bar u_{\xi}(2)u_{\xi}(1)u(1)+\eta_{26}\bar u_{\xi}(2)u(2)u(1)+\eta_{4}u(1)^2\bar u_{\xi\xi\xi}(1)
\nonumber\\
&\quad+{}\eta_{8}u_{\xi\xi}(1)u_{\xi}(1)\bar u(1)+\eta_{20}\bar u(2)u_{\xi\xi}(1)u(1)+\eta_{24}u_{\xi}(2)\bar u_{\xi}(1)u(1)
\nonumber\\
&\quad+{}\eta_{27}u(1)\bar u_{\xi}(1)u(3)+\eta_{28}\bar u(1)u_{\xi}(1)u(3)+\eta_{29}u(1)u_{\xi}(1)\bar u(3)
\nonumber\\
&\quad+{}\eta_{30}u(1)\bar u(2)u(3)+\eta_{31}\bar u(1)u(2)u(3)+\eta_{32}u(1)u(2)\bar u(3)
\nonumber\\
&\quad+{}\eta_{33}|u(1)|^2u_{\xi}(3)+\eta_{34}u(1)^2\bar u_{\xi}(3),
\nonumber\\
f_{3}(4)&\coloneqq\kappa_{1}u(1)|u(1)|^6+\kappa_{2}|u(1)|^2\bar u(1)u_{\xi}(1)^2+\kappa_{3}|u(1)|^2u(1)|u_{\xi}(1)|^2
\label{f34}\\
&\quad+{}\kappa_{4}u(1)^3\bar u_{\xi}(1)^2+\kappa_{5}|u(1)|^4u_{\xi\xi}(1)+\kappa_{6}|u(1)|^2u(1)^2\bar u_{\xi\xi}(1)
\nonumber\\
&\quad+{}\kappa_{7}|u_{\xi}(1)|^2u_{\xi\xi}(1)+\kappa_{8}u_{\xi}(1)^2\bar u_{\xi\xi}(1)+\kappa_{9}u(1)|u_{\xi\xi}(1)|^2+\kappa_{10}\bar u(1)u_{\xi\xi}(1)^2
\nonumber\\
&\quad+{}\kappa_{11}\bar u(1)u_{\xi}(1)u_{\xi\xi\xi}(1)+\kappa_{12}u(1)\bar u_{\xi}(1)u_{\xi\xi\xi}(1)+\kappa_{13}u(1)u_{\xi}(1)\bar u_{\xi\xi\xi}(1)
\nonumber\\
&\quad+{}\kappa_{14}|u(1)|^2u_{\xi\xi\xi\xi}(1)+\kappa_{15}u(1)^2\bar u_{\xi\xi\xi\xi}(1)+\kappa_{16}|u(1)|^2\bar u(1)u(2)^2
\nonumber\\
&\quad+{}\kappa_{17}|u(1)|^2u(1)|u(2)|^2+\kappa_{18}u(1)^3\bar u(2)^2+\kappa_{19}|u(1)|^2\bar u(1)u_{\xi}(1)u(2)
\nonumber\\
&\quad+{}\kappa_{20}|u(1)|^2u(1)\bar u_{\xi}(1)u(2)+\kappa_{21}|u(1)|^2u(1)u_{\xi}(1)\bar u(2)+\kappa_{22}u(1)^3\bar u_{\xi}(1)\bar u(2)
\nonumber\\
&\quad+{}\kappa_{23}\bar u_{\xi}(1)u_{\xi\xi}(1)u(2)+\kappa_{24}u_{\xi}(1)\bar u_{\xi\xi}(1)u(2)+\kappa_{25}u_{\xi}(1)u_{\xi\xi}(1)\bar u(2)
\nonumber\\
&\quad+{}\kappa_{26}u(1)\bar u_{\xi\xi\xi}(1)u(2)+\kappa_{27}\bar u(1)u_{\xi\xi\xi}(1)u(2)+\kappa_{28}u(1)u_{\xi\xi\xi}(1)\bar u(2)
\nonumber\\
&\quad+{}\kappa_{29}\bar u_{\xi\xi}(1)u(2)^2+\kappa_{30}u_{\xi\xi}(1)|u(2)|^2+\kappa_{31}|u(1)|^4u_{\xi}(2)
\nonumber\\
&\quad+{}\kappa_{32}|u(1)|^2u(1)^2\bar u_{\xi}(2)+\kappa_{33}|u_{\xi}(1)|^2u_{\xi}(2)+\kappa_{34}u_{\xi}(1)^2\bar u_{\xi}(2)
\nonumber\\
&\quad+{}\kappa_{35}\bar u(1)u_{\xi\xi}(1)u_{\xi}(2)+\kappa_{36}u(1)\bar u_{\xi\xi}(1)u_{\xi}(2)+\kappa_{37}u(1)u_{\xi\xi}(1)\bar u_{\xi}(2)
\nonumber\\
&\quad+{}\kappa_{38}u(1)\bar u_{\xi}(1)u_{\xi\xi}(2)+\kappa_{39}\bar u(1)u_{\xi}(1)u_{\xi\xi}(2)+\kappa_{40}u(1)u_{\xi}(1)\bar u_{\xi\xi}(2)
\nonumber\\
&\quad+{}\kappa_{41}|u(1)|^2u_{\xi\xi\xi}(2)+\kappa_{42}u(1)^2\bar u_{\xi\xi\xi}(2)+\kappa_{43}\bar u_{\xi}(1)u(2)u_{\xi}(2)
\nonumber\\
&\quad+{}\kappa_{44}u_{\xi}(1)\bar u(2)u_{\xi}(2)+\kappa_{45}u_{\xi}(1)u(2)\bar u_{\xi}(2)+\kappa_{46}u(1)|u_{\xi}(2)|^2+\kappa_{47}\bar u(1)u_{\xi}(2)^2
\nonumber\\
&\quad+{}\kappa_{48}\bar u(1)u(2)u_{\xi\xi}(2)+\kappa_{49}u(1)\bar u(2)u_{\xi\xi}(2)+\kappa_{50}u(1)u(2)\bar u_{\xi\xi}(2)
\nonumber\\
&\quad+{}\kappa_{51}|u(2)|^2u_{\xi}(2)+\kappa_{52}u(2)^2\bar u_{\xi}(2)+\kappa_{53}|u(1)|^4u(3)+\kappa_{54}|u(1)|^2u(1)^2\bar u(3)
\nonumber\\
&\quad+{}\kappa_{55}\bar u(1)u(3)^2+\kappa_{56}u(1)|u(3)|^2+\kappa_{57}|u(2)|^2u(3)+\kappa_{58}u(2)^2\bar u(3)
\nonumber\\
&\quad+{}\kappa_{59}|u_{\xi}(1)|^2u(3)+\kappa_{60}u_{\xi}(1)^2\bar u(3)+\kappa_{61}u(1)\bar u_{\xi\xi}(1)u(3)+\kappa_{62}\bar u(1)u_{\xi\xi}(1)u(3)
\nonumber\\
&\quad+{}\kappa_{63}u(1)u_{\xi\xi}(1)\bar u(3)+\kappa_{64}u(1)\bar u_{\xi}(1)u_{\xi}(3)+\kappa_{65}\bar u(1)u_{\xi}(1)u_{\xi}(3)
\nonumber\\
&\quad+{}\kappa_{66}u(1)u_{\xi}(1)\bar u_{\xi}(3)+\kappa_{67}|u(1)|^2u_{\xi\xi}(3)+\kappa_{68}u(1)^2\bar u_{\xi\xi}(3)+\kappa_{69}u_{\xi}(1)\bar u(2)u(3)
\nonumber\\
&\quad+{}\kappa_{70}\bar u_{\xi}(1)u(2)u(3)+\kappa_{71}u_{\xi}(1)u(2)\bar u(3)+\kappa_{72}\bar u(1)u_{\xi}(2)u(3)+\kappa_{73}u(1)\bar u_{\xi}(2)u(3)
\nonumber\\
&\quad+{}\kappa_{74}u(1)u_{\xi}(2)\bar u(3)+\kappa_{75}u(1)\bar u(2)u_{\xi}(3)+\kappa_{76}\bar u(1)u(2)u_{\xi}(3)+\kappa_{77}u(1)u(2)\bar u_{\xi}(3).
\nonumber
\end{align}
\end{subequations}
If we indicate with $S_{j}$ and $T_{j}$ respectively the real and imaginary parts of $\eta_{j}$, $j=1$, \ldots, $34$, when $\rho_{2}\not=0$, the $A_{4}$-integrability conditions are represented by $48$ real relations whose expressions we leave for a specific Appendix.

Other integrability conditions corresponding to $M_{4}f_{2}\left(3\right)=M_{2}f_{4}\left(3\right)$ ($A_{3}$-integrability conditions) and to $M_{4}f_{2}\left(5\right)=M_{2}f_{4}\left(5\right)$ ($A_{5}$-integrability conditions) in the subspaces with $u\left(2n\right)=0$, $n\geq1$ for purely imaginary coefficients can be found in \cite{S}. They are respectively given by $1$ and $14$ real relations, the first of which can be deduced from (\ref{Siculi}) and corresponds to $I_{4}=\rho_{1}I_{1}/\rho_{2}+\left(I_{2}-I_{3}-2I_{5}\right)/4$.

The results presented in this Section will be used in the following Sections to classify integrable nonlinear equation on the square lattice.

\section{Dispersive affine-linear equations on the square lattice} \label{s2}

The aim of this Section is to derive necessary conditions for the $S$-integrability of  the simplest class of $\mathbb{Z}^2$-lattice equations, that of dispersive and multilinear equations~(\ref{eq}) defined on the square lattice, satisfying the condition~(1) with dispersion relation $\omega_{+}(k)$, i.e.  
\bea
&&\CQ^+\coloneqq a_1 (u_{n,m} + u_{n+1,m+1}) + a_2 (u_{n+1,m} + u_{n,m+1}) \label{q4} \\
 && \qquad +  (\alpha_1-\alpha_2) \, u_{n,m}u_{n+1,m} +   (\alpha_1+\alpha_2)\,
 u_{n,m+1}u_{n+1,m+1} \nonumber \\
&&\qquad  + \,  (\beta_1-\beta_2)\, u_{n,m}u_{n,m+1} + (\beta_1+\beta_2)\,
u_{n+1,m}u_{n+1,m+1} \nonumber \\
&&\qquad + \, \gamma_1 u_{n,m}u_{n+1,m+1} + \gamma_2 u_{n+1,m}u_{n,m+1} \nonumber \\
&&\qquad + \, (\xi_1-\xi_3)\, u_{n,m}u_{n+1,m}u_{n,m+1} + (\xi_1+\xi_3)\,
u_{n,m}u_{n+1,m}u_{n+1,m+1} \nonumber \\
&&\qquad + \, (\xi_2-\xi_4)\, u_{n+1,m}u_{n,m+1}u_{n+1,m+1} + (\xi_2+\xi_4)\,
u_{n,m}u_{n,m+1}u_{n+1,m+1} \nonumber \\
&&\qquad + \, \zeta u_{n,m} u_{n+ 1,m} u_{n,m + 1} u_{n + 1,m + 1}=0, \nonumber
\eea
where $a_1,a_2 \in \mathbb R\setminus \{0\}$, $|a_1| \neq |a_2|$, are the coefficients appearing in the
linear part while  $\alpha_1,\alpha_2,\beta_1,\beta_2,$ $\gamma_1,\gamma_2,$
$\xi_1, \xi_2, \xi_3, \xi_4, \zeta$ are some real parameters which enter in the nonlinear part of the system. Here we will look, by using 
the multiscale procedure described in Section \ref{s1}, into the values of these coefficients for the class $\CQ^+$ to be $A_1$ integrable.


\begin{figure}[htbp]
\begin{center}
\setlength{\unitlength}{0.1em}
\begin{picture}(200,140)(-50,-20)

  \put( 0,  0){\vector(1,0){100}}
  \put( 100,  0){\vector(-1,0){100}}

  \put( 0,100){\vector(1,0){100}}
  \put( 100,100){\vector(-1,0){100}}

  \put(  0, 0){\vector(0,1){100}}
  \put(  0, 100){\vector(0,-1){100}}

  \put(100, 0){\vector(0,1){100}}
  \put(100, 100){\vector(0,-1){100}}

  \put(0, 0){\line(1,1){100}}
  \put(100, 0){\line(-1,1){100}}
   \put(97, -3){$\bullet$}
   \put(-3, -3){$\bullet$}
   \put(-3, 97){$\bullet$}
   \put(97, 97){$\bullet$}
  \put(-32,-13){$u_{n,m}$}
    \put(-45,47){$\beta_1-\beta_2$}
    \put(110,47){$\beta_1+\beta_2$}
     \put(30,-15){$\alpha_1-\alpha_2$}
     \put(30,110){$\alpha_1+\alpha_2$}
     \put(65,80){$\gamma_1$}
     \put(25,80){$\gamma_2$}
  \put(103,-13){$u_{n+1,m}$}
  \put(103,110){$u_{n+1,m+1}$}
  \put(-32,110){$u_{n,m+1}$}
\end{picture}
\caption{Representation of the quadratic nonlinearities of $\CQ_\pm$}
\end{center}
\end{figure}

To perform a classification of the equations $\CQ^+$, we need to find  the  set of transformations that leave it invariant, i.e.~the equivalence transformation. As mentioned before, a generic multilinear equation of the form (\ref{e})  is  invariant under a M\"obius transformation~(\ref{eqMob}). The constant term~$f_0$ and the differences~$a_{00}-a_{11}$, $a_{01}-a_{10}$ transform according to
{\small\bea\label{eqTr3}
&&f_0\overset{T}{\mapsto} f_0'=D^4f_0+
B^4\zeta+
2 B^3D
   \left(\xi _1+\xi
   _2\right)+ B^2 D^2
   \left[\gamma_1+\gamma_2+2
   \left(\alpha _1+\beta
   _1\right)\right]\\ \nonumber
   &&\qquad \qquad+2BD^3
   \left(a_{00}+a_{11}+a_{01}+a_{10}\right),
\\ \nonumber
&&a_{00}-a_{11}\overset{T}{\mapsto}a_{00}'-a_{11}'=(A D-B C) \left[D^2 \left(a_{00}-a_{11}\right)+B^2
   \left(\xi _1-\xi _2-\xi_3+\xi_4\right)-2 B D
   \left(\alpha _2+\beta _2\right)\right]
\\ \nonumber
&&a_{01}-a_{10}\overset{T}{\mapsto} a_{01}'-a_{10}'=(A D-BC) \left[D^2 \left(a_{01}-a_{10}\right)-B^2
   \left(\xi _1-\xi _2+ \xi _3-\xi _4\right)+2 B D
   \left(\alpha _2-\beta _2\right)\right]
\eea}
These formulas allow to determine when a given linear-affine equation~(\ref{e}) can be transformed into one belonging to class~$\CQ^+$. For this to happen all three terms must be null, so setting the l.h.s.~of~(\ref{eqTr3}) to zero we get three polynomial equations over~$B/D$ or~$D/B$. If simultaneously solvable (over the reals), we have an equation of the class~$\CQ^+$. One could try to write the conditions over the coefficients of a general linear-affine equation~(\ref{e}) by using resultant calculations on the three polynomial conditions, but they turn out to be too complicated to merit further attention. 
Thus (\ref{eqTr3})  tells that the class~$\CQ^ +$ is invariant under restricted simultaneous M\"obius transformations $R$ of the form
\begin{equation}
    \label{M}
u_{n,m}\mapsto u_{n,m}'=u_{n,m}/(Cu_{n,m}+D),
\end{equation}
which will be our equivalence transformation. Under ~(\ref{M}) the coefficients of Eq.~(\ref{q4}) undergo the following transformations:
\begin{gather}\nonumber
a_1\overset{R}{\mapsto} a_1'= D^3a_1,\ \ \ a_2\overset{R}{\mapsto} a_2'= D^3a_2,\ \ \ \alpha_1\overset{R}{\mapsto}\alpha_1'= D^2 \left[\alpha _1+C\left(a_1+a_2\right)
   \right],\ \ \ \alpha_2\overset{R}{\mapsto}\alpha_2'= D^2\alpha_2,
\\  \nonumber
\beta_1\overset{R}{\mapsto} \beta_1'= D^2 \left[\beta _1+C\left(a_1+a_2\right)
   \right],\quad \beta_2\overset{R}{\mapsto}\beta_2'= D^2\beta_2,
\\ \nonumber
\gamma_1\overset{R}{\mapsto} \gamma_1'= D^2 \left(\gamma_1+2C a_1 \right),\quad \gamma_2\overset{R}{\mapsto} \gamma_2'= D^2 \left(\gamma_2+2C a_2 \right),
\\ \label{eqtrzeta}
\xi_1\overset{R}{\mapsto} \xi_1'= D \xi _1+\tfrac{1}{2} C D \left[3C
   \left(a_1+a_2\right)
   +\gamma_1+\gamma_2+2 \left(\alpha
   _1-\alpha _2+\beta
   _1\right)\right],
\\ \nonumber
\xi_2\overset{R}{\mapsto} \xi_2'= D \xi _2+\tfrac{1}{2} C D \left[3C
   \left(a_1+a_2\right)
   +\gamma_1+\gamma_2+2 \left(\alpha
   _1+\alpha _2+\beta
   _1\right)\right],
\\ \nonumber
\xi_3\overset{R}{\mapsto} \xi_3'= D \xi _3+\tfrac{1}{2} C D \left[C(a_1
   -a_2)+\gamma_1-\gamma_2+2 \beta
   _2\right],
\\ \nonumber
\xi_4\overset{R}{\mapsto} \xi_4'= D \xi _4+\tfrac{1}{2} C D \left[C(a_1
   -a_2)+\gamma_1-\gamma_2-2 \beta
   _2\right],
\\ 
\zeta\overset{R}{\mapsto} \zeta'= \zeta+C^2 \left[2C
   \left(a_1+a_2\right)
   +\gamma_1+\gamma_2+2 \left(\alpha
   _1+\beta _1\right)\right]+2C
   \left(\xi _1+\xi
   _2\right).\nonumber
\end{gather}
We will indicate by~$\CN$ the number of free parameters (although not all of them essential under $R$) appearing in each subcase of~\fref{q4}. Its maximum number is~$\CN=13$, the number of free coefficients in~\fref{q4}.

\subsection{Classification at order $\ep^3$.}\ \\

By performing the multiscale expansion of Eq.~(\ref{q4}), the following statement holds regarding $A_{1}$-asymptotic integrability

\begin{prop} \label{the1}
The lowest order necessary conditions for the $S$-integrability of equations $\CQ^+$ read:
\begin{itemize}
\item Case 1 ($\CN=9$):
\beq \label{S1}
\left\{ \begin{array}{l}
\alpha_2 = \beta_2 = 0, \\
{\displaystyle{  {\xi_1 = \xi_2}}},\quad{\displaystyle{ {\xi_3 = \xi_4}}}.
\end{array} \right.
\eeq
\item Case 2 ($\CN=7$) :
\beq \label{S2}
\left\{ \begin{array}{l}
\alpha_2 =  \beta_2,\quad \alpha_1 = \beta_1,  \\
a_1 =2 a_2 , \\
\gamma_1 = 2 \gamma_2,\\
a_1 (\xi_1 - \xi_2) = - a_1 (\xi_3 - \xi_4) = -2\alpha_2 \gamma_2.
\end{array} \right.
\eeq
\item Case 3 ($\CN=7$):
\beq \label{S3}
\left\{ \begin{array}{l}
\alpha_2 = - \beta_2,\quad \alpha_1 = \beta_1,  \\
a_2 =2 a_1 , \\
\gamma_2 = 2 \gamma_1,\\
a_1 (\xi_1 - \xi_2) = a_1 (\xi_3 - \xi_4)=
- \alpha_2 \gamma_1.
\end{array} \right.
\eeq
\item Case 4 ($\CN=8$):
\beq \label{S4}
\left\{ \begin{array}{l}
a_2\alpha_1 = a_2\beta_1=\frac12 (a_1+
    a_2)\gamma_2, \\
a_2\gamma_1=a_1\gamma_2,\\
a_1 (\xi_1 - \xi_2) = -\alpha_2 \gamma_1,  \\
a_1 (\xi_3 - \xi_4) = \beta_2 \gamma_1 .
\end{array} \right.
\eeq
\item Case 5 ($\CN=8$):
\beq \label{S5}
\left\{ \begin{array}{l}
(a_2-a_1)\beta_2= (a_2+a_1) \alpha_2,  \\
 2 a_1 a_2 (a_1 - a_2) \alpha_1 = (a_1 + a_2) (\gamma_2 a_1^2-\gamma_1
a_2^2),    \\
2 a_1 a_2 \beta_1 = \gamma_1 a_2^2 + \gamma_2 a_1^2,
 \\
(a_2- a_1)(\xi_1 - \xi_2)= 
(\gamma_1 - \gamma_2)\alpha_2,  \\
(a_2 - a_1)^2(\xi_3 - \xi_4)= \left[ \gamma_2 (a_2 - 3 a_1 )-
\gamma_1 (a_1 - 3 a_2 )\right] \alpha_2  .
\end{array} \right.
\eeq
\item Case 6 ($\CN=8$):
\beq \label{S6}
\left\{ \begin{array}{l}
(a_2+a_1)\beta_2= (a_2- a_1) \alpha_2,  \\
2 a_1 a_2  \alpha_1 = \gamma_1 a_2^2 + \gamma_2 a_1^2,
 \\
 2 a_1 a_2 (a_1 - a_2) \beta_1 =(a_1 + a_2) (\gamma_2 a_1^2-\gamma_1
a_2^2),    \\
(a_2^2 - a_1^2)(\xi_1 - \xi_2)= \left[ \gamma_1 (a_1 - 3 a_2 )-
\gamma_2 (a_2 - 3 a_1 )\right] \alpha_2,    \\
(a_1+ a_2)(\xi_3 - \xi_4)= (\gamma_2 - \gamma_1)\alpha_2.
\end{array} \right.
\eeq
\end{itemize}
The obtained six subclasses of equation~\fref{q4} are invariant under the restricted  M\"obius transformation~(\ref{M}).
\end{prop}

{\underline{Proof:}} Following the procedure described in Section \ref{s1} we 
expand the fields appearing in equation $\CQ^+$ according to formulas
(\ref{bas}-\ref{nm}).
The lowest order necessary conditions for the $S$-integrability of $\CQ^+$ are obtained
by considering the equation $\CW_3$ (see Eq.~(\ref{eee})), namely the order
$\ep^3$ of the multiscale expansion. At this order we get
the $m_2$-evolution equation for the harmonic $u_0^{(1)}$, that is a \emph{NLS}
equation of the form
\beq
\ri \delta_{m_2} u_1^{(1)} + \rho_1\delta_{\xi}^2 u_1^{(1)}+ 
 \rho_2 u_1^{(1)} |u_1^{(1)}|^2=0, \qquad \xi\coloneqq n_1 - \frac{d\omega}{d \kappa} m_1, \label{nls}
\eeq
where the coefficients $\rho_1$ and $\rho_2$ will depend on the parameters of 
the equation $\CQ^+$ and on the wave parameters $\kappa$ and $\omega=\omega_+$, with
$\omega_+$ expressed in terms of $\kappa$ through the dispersion relation~(\ref{e5+}). 
According to our multiscale test the lowest order necessary
condition for $\CQ^+$ to be an $S$-integrable lattice equation is that Eq.~(\ref{nls}) be integrable itself, namely $\rho_1$ and $\rho_2$ have
to be real coefficients.

Let us outline the construction of Eq.~(\ref{nls}).
At $\CO(\ep)$ we get:
\begin{itemize}
\item for $\alpha=1$ a linear equation which is identically satisfied by the
  dispersion relation (\ref{e5+}).
\item for $\alpha=0$ a linear equation whose solution is $u_1^{(0)}=0$.
\end{itemize}
At $\CO(\ep^2)$, taking into account the dispersion relation (\ref{e5+}),
we get:
\begin{itemize}
\item for $\alpha=2$ an algebraic relation between $u_2^{(2)}$ and $u_1^{(1)}$.
\item for $\alpha=1$ a linear wave equation for $u_1^{(1)}$, whose solution is
  given by $u_1^{(1)}(n_1,m_1,m_2)=u_1^{(1)}(\xi,m_2)$, where 
$\xi\coloneqq n_1 - (d\omega/d \kappa) m_1$.
\item for $\alpha=0$ an algebraic relation between $u_2^{(0)}$ and $u_1^{(1)}$.
\end{itemize}
Notice that from the $\CO(\ep^2)$ we find that the dependence of all the harmonics
on the slow-variables $n_1$ and $m_1$ is given by $\xi$.

At $\CO(\ep^3)$, for $\alpha=1$, by using the results obtained at the previous
orders, one gets the \emph{NLS} equation (\ref{nls}) with
$$
\rho_1 = \frac{a_1 a_2 (a_1^2-a_2^2) \sin \kappa }{ (a_1^2+a_2^2+2 a_1 a_2
  \cos \kappa)^2}, \qquad 
\rho_2 = \CR_1 + \ri \CR_2, \label{r2} 
$$
where
\bea
&& \CR_1=\frac{\sin \kappa \left[ \CR_1^{(0)} + \CR_1^{(1)} \cos \kappa 
+ \CR_1^{(2)} \cos^2 \kappa+ \CR_1^{(3)} \cos^3 \kappa +\CR_1^{(4)} \cos^4 \kappa  \right]}
{(a_1+a_2)(a_1^2+a_2^2+2 a_1 a_2 \cos \kappa)^2 \left[(a_1-a_2)^2 + 2 a_1 a_2 \cos
  \kappa (1+ \cos \kappa) \right]}, \label{r21} \\
&& \CR_2=\frac{\CR_2^{(0)} + \CR_2^{(1)} \cos \kappa 
+ \CR_2^{(2)} \cos^2 \kappa+ \CR_2^{(3)} \cos^3 \kappa +\CR_2^{(4)} \cos^4
\kappa
+\CR_2^{(5)} \cos^5 \kappa}
{(a_1+a_2)(a_1^2+a_2^2+2 a_1 a_2 \cos \kappa)^2 \left[(a_1-a_2)^2 + 2 a_1 a_2 \cos
  \kappa (1+ \cos \kappa) \right]}. \label{r22}
\eea
Here the coefficients $\CR_1^{(i)}$, $0 \leq i \leq 4$, and $\CR_2^{(i)}$, $0
\leq i \leq 5$, are polynomials 
depending on the coefficients $a_1,a_2 ,\alpha_1,\alpha_2,\beta_1,\beta_2,$ $\gamma_1,\gamma_2,$
$\xi_1,...,\xi_4$ and their expressions are cumbersome, so that we omit
them.

Note that $\rho_1$ is a real coefficient depending only on the parameters
of the linear part of $\CQ^+$, while $\rho_2$ is a complex one. Hence
the integrability of the \emph{NLS} equation (\ref{nls}) is equivalent to the request
$\CR_2 =0 \; \forall \, \kappa$, that is 
\beq
\CR_2^{(i)}=0, \qquad 0\leq i \leq 5. \label{sys}
\eeq
Eq.~(\ref{sys}) is a nonlinear algebraic system of six equations in twelve unknowns.  By solving it  one gets the six solutions contained in Proposition 1. 
These solutions are computed taking into account that 
$a_1,a_2 \in \mathbb R\setminus \{0\}$ with $|a_1| \neq |a_2|$.
One can solve two of the  six equations (\ref{sys}) for $\xi_1$ 
and $\xi_3$, thus expressing them in terms of the remaining ten coefficients.
The resulting system of four equations turns out to be
$\xi_2$ and $\xi_4$-independent and linear in the 
four variables~$\alpha_1$, $\beta_1$, $\gamma_1$ and $\gamma_2$. Therefore we may write the remaining four equations as 
a matrix equation with coefficients nonlinearly depending on 
$\alpha_2$, $\beta_2$, $a_1$ and $a_2$. The rank of the matrix is three. The six solutions are obtained by requiring that the matrix be of rank 3, 2, 1  and 0, and correspond to six classes of equations~\fref{q4} that pass integrability conditions up to order~$ \mathcal{O}(\varepsilon^3)$. A direct calculation proves the invariance of the six classes with respect to the restricted M\"obius transformation $R$.
\endpf
\begin{cor}
If the coefficients $a_1,a_2 ,\alpha_1,\alpha_2,\beta_1,\beta_2,$ $\gamma_1,\gamma_2,$
$\xi_1,...,\xi_4$ of equation $\CQ^+$ do not satisfy one of the conditions
given in Eqs.~(\ref{S1}--\ref{S6}) then $\CQ^+$ is not integrable.
\end{cor}

 \emph{Quadratic difference equations} are a subclass of $\CQ^+$ which have attracted a deal of attention. These equations are not M\"obius invariant, but we can spot those that belong to the class~$\CQ^+$ and pass our integrability conditions, just by inspection of~(\ref{S1}--\ref{S4}). 

\subsection{Classification at order $\ep^4$.}\ \\

\indent For what concerns the $A_{2}$-asymptotically integrable cases satisfying the integrability conditions (\ref{CieloUrbico}), the following statement holds
\begin{prop}\label{the11}
At order $\ep^4$, the necessary conditions for the $S$-integrability of equations $\CQ^+$ read:
\begin{itemize}
\item Case 1 ($\CN=9$):
\beq \label{S11}
\left\{ \begin{array}{l}
\alpha_2 = \beta_2 = 0, \\
{\displaystyle{  {\xi_1 = \xi_2}}},\quad{\displaystyle{ {\xi_3 = \xi_4}}}.
\end{array} \right.
\eeq
\item Case 4 ($\CN=8$):
\beq \label{S44}
\left\{ \begin{array}{l}
\alpha_{1}=\beta_{1}=\frac{\left(a_{1}+a_{2}\right)\gamma_{1}}{2a_{1}},\\
\gamma_{2}=\frac{a_{2}\gamma_{1}}{a_{1}},\\
a_1 (\xi_1 - \xi_2) = -\alpha_2 \gamma_1, \\
a_1 (\xi_3 - \xi_4) = \beta_2 \gamma_1 ,\\
\left(\alpha_{2},\beta_{2}\right)\not=\left(0,0\right).
\end{array} \right.
\eeq
\end{itemize}
\vspace{2mm}
The corresponding two subclasses of equations are non overlapping and invariant under the restricted M\"obius transformation~(\ref{M}).
\end{prop}
Notice that of the six $A_{1}$-asymptotically integrable cases listed in Proposition~\ref{the1}, Case~\emph{1} and Case~\emph{4} automatically satisfy the $A_{2}$-integrability conditions (\ref{CieloUrbico}), while the remaining four cases \emph{2}, \emph{3}, \emph{5} and \emph{6} specify to some subcases of theirs.
Notice that only two out the previous four quadratic cases in {\bf Remark 1} survive, the Cases Q1 and Q4: the first one is a subcase of Case 1, while the second is a subcase of Case 4.

\subsection{Classification at order $\ep^5$.}\ \\

\indent It is possible to find all the cases satisfying the $A_{3}$-integrability conditions (\ref{Siculi}). They are given by the following proposition    
\begin{prop}
The necessary and sufficient conditions for $\ep^{5}$ asymptotic integrability are:\\\\\indent Case (a): ($\CN=4$)
\bea
\nonumber\alpha_{2}=\beta_{2}=0,\ \ \ \gamma_{2}=\alpha_{1}+\beta_{1}-\gamma_{1},\ \ \ a_{2}=2a_{1},\ \ \ \left(2\alpha_{1}-3\gamma_{1},\ 2\beta_{1}-3\gamma_{1}\right)\not=\left(0,\ 0\right),\\
\nonumber\xi_{1}=\xi_{2}=\frac{\alpha_{1}\beta_{1}} {2a_{1}},\ \ \ \xi_{3}=\xi_{4}=-\frac{\left(\alpha_{1}-\gamma_{1}\right)\left(\beta_{1}-\gamma_{1}\right)} {2a_{1}},\ \ \ \zeta=\frac{\gamma_{1}\left[3\gamma_{1}^2-3\gamma_{1}\left(\alpha_{1}+\beta_{1}\right)+4\alpha_{1}\beta_{1}\right]} {4a_{1}^2};
\eea
\indent Case (b): ($\CN=4$)
\bea
\nonumber\alpha_{2}=\beta_{2}=0,\ \ \ \gamma_{1}=\alpha_{1}+\beta_{1}-\gamma_{2},\ \ \ a_{1}=2a_{2},\ \ \ \left(2\alpha_{1}-3\gamma_{2},\ 2\beta_{1}-3\gamma_{2}\right)\not=\left(0,\ 0\right)\\
\nonumber\xi_{1}=\xi_{2}=\frac{\alpha_{1}\beta_{1}} {2a_{2}},\ \ \ \xi_{3}=\xi_{4}=\frac{\left(\alpha_{1}-\gamma_{2}\right)\left(\beta_{1}-\gamma_{2}\right)} {2a_{2}},\ \ \ \zeta=\frac{\gamma_{2}\left[3\gamma_{2}^2-3\gamma_{2}\left(\alpha_{1}+\beta_{1}\right)+4\alpha_{1}\beta_{1}\right]} {4a_{2}^2};
\eea
\indent Case (c): ($\CN=5$)
\bea
\nonumber\alpha_{1}=\beta_{1}=\frac{\left(a_{1}+a_{2}\right)\gamma_{1}} {2a_{1}},\ \ \ \alpha_{2}=\beta_{2}=0,\ \ \ \gamma_{2}=\frac{a_{2}\gamma_{1}} {a_{1}},\ \ \ \xi_{1}=\xi_{2},\\
\nonumber\xi_{3}=\xi_{4}=\frac{\left(a_{2}-a_{1}\right)\gamma_{1}^2}{4a_{1}^2}-\frac{\left(a_{2}-a_{1}\right)} {\left(a_{2}+a_{1}\right)}\xi_{2},\ \ \ \rho\coloneqq\left[\frac{8a_{1}^2\xi_{2}} {\left(a_{1}+a_{2}\right)}-3\gamma_{1}^2\right]\frac{1}{\left(a_{1}+a_{2}\right)^2}\not=0;
\eea
\indent Case (d): ($\CN=5$)
\bea
\nonumber\alpha_{1}=\beta_{1}=\frac{\left(a_{1}+a_{2}\right)\gamma_{1}} {2a_{1}},\ \ \ \alpha_{2}=\beta_{2}=0,\ \ \ \gamma_{2}=\frac{a_{2}\gamma_{1}} {a_{1}},\ \ \ \xi_{1}=\xi_{2},\\
\nonumber\xi_{3}=\xi_{4}=\frac{\left(a_{1}-a_{2}\right)\gamma_{1}^2}{2a_{1}^2}-\frac{\left(a_{1}-a_{2}\right)} {\left(a_{1}+a_{2}\right)}\xi_{2},\ \ \ \rho\coloneqq\left[\frac{8a_{1}^2\xi_{2}} {\left(a_{1}+a_{2}\right)}-3\gamma_{1}^2\right]\frac{1}{\left(a_{1}+a_{2}\right)^2}\not=0;
\eea
\indent Case (e): ($\CN=4$)
\bea
\nonumber\alpha_{1}=\beta_{1}=\frac{\gamma_{1}+\gamma_{2}} {2},\ \ \ \alpha_{2}=\beta_{2}=0,\ \ \ \gamma_{2}\not=\frac{a_{2}\gamma_{1}}{a_{1}},\ \ \ \frac{a_{2}}{a_{1}}\not=\frac{1}{2},\ 2,\\
\nonumber\xi_{1}=\xi_{2}=\frac{3\left(\gamma_{1}+\gamma_{2}\right)^2} {8\left(a_{1}+a_{2}\right)},\ \ \ \xi_{3}=\xi_{4}=\frac{9\left(a_{1}-a_{2}\right)\left(a_{1}\gamma_{2}-a_{2}\gamma_{1}\right)^2}{8a_{1}a_{2}\left(a_{1}+a_{2}\right)^2}-\frac{a_{1}\gamma_{2}^2-a_{2}\gamma_{1}^2}{8a_{1}a_{2}},\\
\nonumber\zeta=\frac{\left(\gamma_{1}+\gamma_{2}\right)^3}{4\left(a_{1}+a_{2}\right)^2}+\frac{\left(a_{1}-a_{2}\right)\left(a_{1}\gamma_{2}-a_{2}\gamma_{1}\right)^3}{a_{1}^2a_{2}^2\left(a_{1}+a_{2}\right)^2};
\eea
\end{prop}

\indent{\bf Notes:} In all of the cases $a_{2}/a_{1}\not=(0$, $\pm 1)$; the values $a_{2}/a_{1}=(2, \,\frac12)$ are excluded in Case~(e) because we would obtain respectively a subcase of Case (a) or of Case (b). All of the Cases~(a)--(e) are subcases of Case~1. So nothing survives out of Case~4 at order $\ep^5$. 
 Cases Q$_{\alpha}$-Q$_{\delta}$ are subcases both of the Case Q1 and Case (a); the Cases Q$_{\eta}$-Q$_{\lambda}$ are subcases both of the Case Q1 and Case (b).

\subsubsection{Canonical forms for $\varepsilon^5$ $S$-asymptotically integrable cases. Comparison with the ABS list.}
We will use now the M\"obius transformation to reduce the equation to normal form, i.e. to eliminate the maximum number of free parameters appearing in the nonlinear difference equation and reduce the coefficients of the linear part in $v_{n,m}$ and $v_{n+1,m+1}$ to 1.

In the Case (a) of Proposition 4, performing the M\"obius transformation
\bea
\nonumber u_{n,m}=\frac{\alpha v_{n,m}+\beta} {\gamma v_{n,m}+\delta},
\eea
with
\bea
\nonumber \beta=0,\ \ \ \ \gamma=-\frac{\gamma_{1}\delta} {2},\ \ \ \ \alpha=a_{1}\delta,\ \ \ \delta\not=0,
\eea
we obtain the canonical form:\\\\
\indent \emph{Case (a$^{\prime}$): ($\CN=2$)}
\bea
\nonumber &&v_{n,m}+v_{n+1,m+1}+2\left(v_{n+1,m}+v_{n,m+1}\right)+v_{n+1,m}v_{n,m+1}\left(\tau_{1}+\tau_{2}\right)\\
\nonumber &&\quad+\left(v_{n+1,m}v_{n+1,m+1}+v_{n,m}v_{n,m+1}\right)\tau_{2}+\left(v_{n,m+1}v_{n+1,m+1}+v_{n,m}v_{n+1,m}\right)\tau_{1}\\
\label{s1a} &&\quad+v_{n+1,m}v_{n,m+1}\left(v_{n,m}+v_{n+1,m+1}\right)\tau_{1}\tau_{2}=0,
\eea
where $(\tau_{1}, \tau_{2})\coloneqq\left(\alpha_{1}-\frac{3\gamma_{1}}{2}, \beta_{1}-\frac{3\gamma_{1}}{2}\right)\not=(0, 0)$. Performing a further rescaling  on (\ref{s1a}), we can fix, in all generality, the coefficients to either  $\tau_{1}=0$ and $\tau_{2}=1$ or $\tau_{1}=1$ and we obtain  the following two canonical forms respectively
\begin{subequations}
\bea
&&v_{n,m}+v_{n+1,m+1}+2\left(v_{n+1,m}+v_{n,m+1}\right)+\label{Lebano1}\\
\nonumber &&\quad+v_{n+1,m}v_{n,m+1}+v_{n+1,m}v_{n+1,m+1}+v_{n,m}v_{n,m+1}=0,\\
&&v_{n,m}+v_{n+1,m+1}+2\left(v_{n+1,m}+v_{n,m+1}\right)+v_{n+1,m}v_{n,m+1}\left(1+\tau_{2}\right)+\label{Lebano2}\\
\nonumber &&\quad+\left(v_{n+1,m}v_{n+1,m+1}+v_{n,m}v_{n,m+1}\right)\tau_{2}+v_{n,m+1}v_{n+1,m+1}+v_{n,m}v_{n+1,m}\\
\nonumber &&\quad+v_{n+1,m}v_{n,m+1}\left(v_{n,m}+v_{n+1,m+1}\right)\tau_{2}=0,
\eea
\end{subequations}
representing the two non overlapping subclasses of \emph{Case (a)} defined respectively by the additional conditions $\alpha_{1}=\frac{3\gamma_{1}}{2}$ and $\alpha_{1}\not=\frac{3\gamma_{1}}{2}$. As under a restricted M\"obious transformation $\tau_{2}$ is invariant, we see that two canonical forms (\ref{Lebano2}), specified by two invariants $\tau_{2a}$ and $\tau_{2b}$, form two disconnected components of the same conjugacy subclass unless $\tau_{2a}=\tau_{2b}$;\\\\ 
\indent In the Case (b) of Proposition 4, performing the M\"obius transformation
\bea
\nonumber u_{n,m}=\frac{\alpha v_{n,m}+\beta} {\gamma v_{n,m}+\delta},
\eea
with
\bea
\nonumber \beta=0,\ \ \ \ \gamma=-\frac{\gamma_{2}\delta} {2},\ \ \ \ \alpha=a_{2}\delta,\ \ \ \delta\not=0,
\eea
we obtain the canonical form:\\\\
\indent \emph{Case (b$^{\prime}$): ($\CN=2$)}
\bea
\nonumber &&2\left(v_{n,m}+v_{n+1,m+1}\right)+v_{n+1,m}+v_{n,m+1}+v_{n,m}v_{n+1,m+1}\left(\tau_{1}+\tau_{2}\right)\\
\nonumber &&\quad+\left(v_{n+1,m}v_{n+1,m+1}+v_{n,m}v_{n,m+1}\right)\tau_{2}+\left(v_{n,m+1}v_{n+1,m+1}+v_{n,m}v_{n+1,m}\right)\tau_{1}\\
\label{s2a} &&\quad+v_{n,m}v_{n+1,m+1}\left(v_{n+1,m}+v_{n,m+1}\right)\tau_{1}\tau_{2}=0,
\eea
where $(\tau_{1}, \tau_{2})\coloneqq\left(\alpha_{1}-\frac{3\gamma_{2}}{2}, \beta_{1}-\frac{3\gamma_{2}}{2}\right)\not=(0, 0)$. Performing a further rescaling on (\ref{s2a}) we can fix, in all generality, the parameters either to $\tau_{1}=0$ and $\tau_{2}=1$ or to $\tau_{1}=1$ and  we obtain respectively the two canonical forms
\begin{subequations}
\bea
&&2\left(v_{n,m}+v_{n+1,m+1}\right)+v_{n+1,m}+v_{n,m+1}+\label{Lebano3}\\
\nonumber &&\quad+v_{n,m}v_{n+1,m+1}+v_{n+1,m}v_{n+1,m+1}+v_{n,m}v_{n,m+1}=0,\\
&&2\left(v_{n,m}+v_{n+1,m+1}\right)+v_{n+1,m}+v_{n,m+1}+v_{n,m}v_{n+1,m+1}\left(1+\tau_{2}\right)+\label{Lebano4}\\
\nonumber &&\quad+\left(v_{n+1,m}v_{n+1,m+1}+v_{n,m}v_{n,m+1}\right)\tau_{2}+v_{n,m+1}v_{n+1,m+1}+v_{n,m}v_{n+1,m}\\
\nonumber &&\quad+v_{n,m}v_{n+1,m+1}\left(v_{n+1,m}+v_{n,m+1}\right)\tau_{2}=0,
\eea
\end{subequations}
representing the two non overlapping subclasses of \emph{Case (b)} defined respectively by the additional conditions $\alpha_{1}=\frac{3\gamma_{2}}{2}$ and $\alpha_{1}\not=\frac{3\gamma_{2}}{2}$. As  $\tau_{2}$ is invariant under a restricted M\"obious transformation, we see that two canonical forms (\ref{Lebano4}), specified by two invariants $\tau_{2a}$ and $\tau_{2b}$, form two disconnected components of the same conjugacy subclass unless $\tau_{2a}=\tau_{2b}$;\\\\
\indent In the Cases (c) and (d) of Proposition 4, performing the  M\"obius transformation
\bea
\nonumber u_{n,m}=\frac{\alpha v_{n,m}+\beta} {\gamma v_{n,m}+\delta},
\eea
with
\bea
\nonumber \alpha=\frac{2a_{1}\delta} {\left(a_{1}+a_{2}\right)\sqrt{\left|\rho\right|}},\ \ \ \beta=0,\ \ \ \gamma=-\frac{\gamma_{1}\delta} {\left(a_{1}+a_{2}\right)\sqrt{\left|\rho\right|}},\ \ \ \delta\not=0,
\eea
we obtain the canonical forms:

 \emph{Case (c$^{\prime}$): ($\CN=2$)}
\bea
&&v_{n,m}+v_{n+1,m+1}+\epsilon\left(v_{n+1,m}+v_{n,m+1}\right)+\label{Kroton1}\\
\nonumber &&\quad+sgn\left(\rho\right)\left[\epsilon v_{n+1,m}v_{n,m+1}\left(v_{n,m}+v_{n+1,m+1}\right)+v_{n,m}v_{n+1,m+1}\left(v_{n+1,m}+v_{n,m+1}\right)\right]\\
\nonumber &&\quad+\zeta^{\prime}v_{n,m}v_{n+1,m}v_{n,m+1}v_{n+1,m+1}=0,
\eea
and

\emph{Case (d$^{\prime}$): ($\CN=2$)}
\bea
&&v_{n,m}+v_{n+1,m+1}+\epsilon\left(v_{n+1,m}+v_{n,m+1}\right)+\label{Kroton2}\\
\nonumber &&\quad+sgn\left(\rho\right)\left[v_{n+1,m}v_{n,m+1}\left(v_{n,m}+v_{n+1,m+1}\right)+\epsilon v_{n,m}v_{n+1,m+1}\left(v_{n+1,m}+v_{n,m+1}\right)\right]\\
\nonumber &&\quad+\zeta^{\prime}v_{n,m}v_{n+1,m}v_{n,m+1}v_{n+1,m+1}=0,
\eea
where $\epsilon\coloneqq a_{2}/a_{1}\not=0,\pm 1$ and $\zeta^{\prime}\coloneqq 8s\left|\frac{\pi^2}{\rho^3}\right|^{1/2}/\left(1+\epsilon\right)^2$, $\pi\coloneqq\left[\zeta-2\frac{\gamma_{1}} {a_{1}}\xi_{2}+\frac{\left(a_{1}+a_{2}\right)\gamma_{1}^3} {2a_{1}^3}\right]/\left(a_{1}+a_{2}\right)$ and $s\coloneqq\pm 1$. As under a restricted M\"obius transformation $\rho\rightarrow\rho\left(\alpha/\delta\right)^2$ and $\pi\rightarrow\pi\left(\alpha/\delta\right)^3$, we see that the absolute value of $\zeta^{\prime}$ and $sgn\left(\rho\right)$ are invariant under such a transformation.  With another rescaling we can always fix $\zeta^{\prime}\geq 0$ and the two canonical forms, specified by the two set of invariants $\left(\epsilon_{a}, sgn\left(\rho_{a}\right), \zeta^{\prime}_{a}\right)$ and $\left(\epsilon_{b}, sgn\left(\rho_{b}\right), \zeta^{\prime}_{b}\right)$, form two disconnected components of the conjugacy class unless the two sets are  the same;

In the Case (e) of Proposition 4, performing the M\"obius transformation
\bea
\nonumber u_{n,m}=\frac{\alpha v_{n,m}+\beta} {\gamma v_{n,m}+\delta},
\eea
with
\bea
\nonumber \beta=0,\ \ \ \ \gamma=-\frac{\left(\gamma_{1}+\gamma_{2}\right)\alpha} {2\left(a_{1}+a_{2}\right)},\ \ \ \ \delta=\frac{\left(a_{2}\gamma_{1}-a_{1}\gamma_{2}\right)\alpha} {a_{1}\left(a_{1}+a_{2}\right)},\ \ \ \alpha\not=0,
\eea
we obtain the canonical form:

\emph{Case (e$^{\prime}$): ($\CN=1$)}
\bea
&&v_{n,m}+v_{n+1,m+1}+\epsilon\left(v_{n+1,m}+v_{n,m+1}\right)+v_{n,m}v_{n+1,m+1}-v_{n+1,m}v_{n,m+1}+\label{Kroton3}\\
\nonumber &&\quad+\left(1-\frac{1} {\epsilon}\right)\left[v_{n+1,m}v_{n,m+1}\left(v_{n,m}+v_{n+1,m+1}\right)-v_{n,m}v_{n+1,m+1}\left(v_{n+1,m}+v_{n,m+1}\right)\right]\\
\nonumber &&\quad+\left(1-\frac{1} {\epsilon^2}\right)v_{n,m}v_{n+1,m}v_{n,m+1}v_{n+1,m+1}=0,
\eea
where $\epsilon\coloneqq a_{2}/a_{1}\not=0,\pm 1,2,1/2$. As $\epsilon$ is invariant under a restricted M\"obius transformation, we see that two canonical forms, specified by the two invariants $\epsilon_{a}$ and $\epsilon_{b}$, form two disconnected components of the conjugacy class unless $\epsilon_{a}=\epsilon_{b}$;



As our allowed transformations are subcases of the full M\"obius transformations allowed in the ABS approach \cite{abs1}, any conjugacy class of ours is either completely contained into one of the ABS classification or is totally disjointed from them. Considering that no one out of the (left hand members of the) canonical forms (a$^{\prime}$)-(e$^{\prime}$) possesses the invariance (up to an overall sign) under $v_{n,m}\leftrightarrow v_{n+1,m}$, $v_{n,m+1}\leftrightarrow v_{n+1,m+1}$, we can conclude that no intersection can exist between our classes and those generated by the $ABS$ list. Even more, no equation in our list is of Klein-type or, that is the same \cite{LY2}, a subcase of the $Q_{V}$ equation.    

We can  enlarge our class of transformations by including also an exchange $n\leftrightarrow m$ between the two independent variables. The subclass (\ref{Lebano1}) can be discarded because under this exchange we would get it from subclass (\ref{Lebano2}) with $\tau_{2}=0$; similarly the subclass (\ref{Lebano3}) can be discarded because under this exchange we would get it from subclass (\ref{Lebano4}) with $\tau_{2}=0$; finally the subclasses (\ref{Kroton1}-\ref{Kroton3}) are invariant under this transformation.

Let us include also  the inversion $n\rightarrow -n$. Setting $\tilde v_{n,m}\coloneqq v_{-n,m}$, we have that, if $v_{n,m}$ satisfies (\ref{Lebano2}), then $\tilde v_{n,m}$ satisfies (\ref{Lebano4}); if $v_{n,m}$ satisfies (\ref{Kroton1}) with parameters $\epsilon$ and $\zeta^{\prime}$, then $\tilde v_{n,m}\coloneqq sgn\left(\epsilon\right)v_{-n,m}$ satisfies (\ref{Kroton1}) with parameters $1/\epsilon$ and $\zeta^{\prime}/\left|\epsilon\right|$ and similarly for Eq.~(\ref{Kroton2}); if $v_{n,m}$ satisfies (\ref{Kroton3}) with parameter $\epsilon$, then $\tilde v_{n,m}\coloneqq -v_{-n,m}/\epsilon$ satisfies (\ref{Kroton3}) with parameter $1/\epsilon$ (this implies that, if $v_{n,m}$ satisfies one of the four canonical forms (\ref{Lebano4}), (\ref{Kroton1}-\ref{Kroton3}), then also $\tilde v_{n,m}\coloneqq v_{-n,-m}$ does). As a consequence under this enlarged class of transformations the Eq.~(\ref{Lebano4}) can be discarded and in the case of the Eqs.~(\ref{Kroton1}-\ref{Kroton3}) we can limit the parameter $\epsilon$ to the range $-1<\epsilon<1$, $\epsilon\not=0$ as the equation with parameters $1/\epsilon$ and $\zeta^{\prime}$ can be obtained from the corresponding with parameters $\epsilon$ and $\zeta^{\prime}\left|\epsilon\right|$.\\

{\bf Notes:} In the Cases (c) and (d) of Proposition 4, when $\pi=0$, corresponding to $\zeta^{\prime}=0$ in the cases (c$^{\prime}$) and (d$^{\prime}$), they reduce to the $S$-integrable cases analyzed in Levi-Yamilov and Ramani-Grammaticos \cite{SGR}.

\subsection{Classification at order $\ep^6$.}\ \\

Now we perform a multiscale reduction at order $\ep^6$ on the four canonical forms (\ref{Lebano4}), (\ref{Kroton1}-\ref{Kroton3}) and  we find that all the so far obtained equations satisfy the $A_{4}$-integrability conditions (\ref{Phoenix}). Hence we can state the following proposition    
\begin{prop}
Up to a restricted M\"obius transformations $\tilde v_{n,m}\coloneqq v_{n,m}/\left(\alpha v_{n,m}+\beta\right)$, exchanges $n\leftrightarrow m$ and inversions $n\rightarrow -n$, all the $A_{4}$-asymptotically $S$-integrable cases in the class ${\mathcal Q}_{+}$ are given by
\begin{subequations}\label{Quadrelli}
\bea
&&v_{n,m}+v_{n+1,m+1}+2\left(v_{n+1,m}+v_{n,m+1}\right)+v_{n+1,m}v_{n,m+1}\left(1+\tau\right)+\label{Quadrelli1}\\
\nonumber &&\quad +\left(v_{n+1,m}v_{n+1,m+1}+v_{n,m}v_{n,m+1}\right)\tau+v_{n,m+1}v_{n+1,m+1}+v_{n,m}v_{n+1,m}\\
\nonumber && \quad +v_{n+1,m}v_{n,m+1}\left(v_{n,m}+v_{n+1,m+1}\right)\tau=0; \\
&&v_{n,m}+v_{n+1,m+1}+\epsilon\left(v_{n+1,m}+v_{n,m+1}\right)+\label{Quadrelli2}\\
\nonumber &&\quad +\delta\left[\epsilon v_{n+1,m}v_{n,m+1}\left(v_{n,m}+v_{n+1,m+1}\right)+v_{n,m}v_{n+1,m+1}\left(v_{n+1,m}+v_{n,m+1}\right)\right]\\
\nonumber && \quad +\tau v_{n,m}v_{n+1,m}v_{n,m+1}v_{n+1,m+1}=0,\ \ \ -1<\epsilon<1,\ \ \ \epsilon\not=0,\ \ \ \delta\coloneqq\pm 1,\ \ \ \tau\geq 0;\\
&&v_{n,m}+v_{n+1,m+1}+\epsilon\left(v_{n+1,m}+v_{n,m+1}\right)+\label{Quadrelli3}\\
\nonumber &&\quad +\delta\left[v_{n+1,m}v_{n,m+1}\left(v_{n,m}+v_{n+1,m+1}\right)+\epsilon v_{n,m}v_{n+1,m+1}\left(v_{n+1,m}+v_{n,m+1}\right)\right]\\
\nonumber &&\quad +\tau v_{n,m}v_{n+1,m}v_{n,m+1}v_{n+1,m+1}=0,\ \ \ -1<\epsilon<1,\ \ \ \epsilon\not=0,\ \ \ \delta\coloneqq\pm 1,\ \ \ \tau\geq 0;\\
&&v_{n,m}+v_{n+1,m+1}+\epsilon\left(v_{n+1,m}+v_{n,m+1}\right)+v_{n,m}v_{n+1,m+1}-v_{n+1,m}v_{n,m+1}+\label{Quadrelli4}\\
\nonumber &&\quad +\left(1-\tfrac{1} {\epsilon}\right)\left[v_{n+1,m}v_{n,m+1}\left(v_{n,m}+v_{n+1,m+1}\right)-v_{n,m}v_{n+1,m+1}\left(v_{n+1,m}+v_{n,m+1}\right)\right]\\
\nonumber &&\quad +\left(1-\tfrac{1} {\epsilon^2}\right)v_{n,m}v_{n+1,m}v_{n,m+1}v_{n+1,m+1}=0,\ \ \ -1<\epsilon<1,\ \ \ \epsilon\not=0,\ \frac{1}{2}.
\eea
\end{subequations}
\end{prop}
Eqs.~(\ref{Quadrelli1}, \ref{Quadrelli4}) depend on $\CN=1$ free parameter, while (\ref{Quadrelli2}, \ref{Quadrelli3}) depend on $\CN=2$ free parameters (without considering the additional discrete parameter $\delta$).

\indent If in (\ref{Quadrelli1}), when $\tau=0$, we apply the (not allowed) transformation $v_{n,m}\coloneqq\sqrt{3}w_{n,m}-1$, we obtain
\begin{equation}
\label{kj1}w_{n,m}w_{n+1,m}+w_{n+1,m}w_{n,m+1}+w_{n,m+1}w_{n+1,m+1}-1=0,
\end{equation}
which in the direction $n$ satisfies two first order necessary integrability conditions given in \cite{LY2} but doesn't admit the corresponding three-point generalized symmetry either autonomous or not, while in the direction $m$ the first order integrability conditions are not satisfied. Following \cite{GH} we were able to prove the integrability of~\eqref{kj1} constructing two five-point symmetries, one in the $n$ direction depending on the points $\left(n+2,m\right)$, $\left(n+1,m\right)$, $\left(n,m\right)$, $\left(n-1,m\right)$ and $\left(n-2,m\right)$ and the other one in the $m$ direction. In \cite{SHL} its integrability was finally proven providing a $3\times 3$ Lax pair. Moreover this equation has the singularity confinement property, can be bilinearized, possesses multisoliton solutions and has a continuous limit into the \emph{mKdV} equation, \cite{GRSS}.\\
\indent If in (\ref{Quadrelli1}), when $\tau=1$, we apply the (not allowed) transformation $v_{n,m}\coloneqq-\left(2^{1/3}w_{n,m}+1\right)$, we obtain
\begin{equation}
\label{jk2}w_{n+1,m}w_{n,m+1}\left(w_{n,m}+w_{n+1,m+1}\right)+1=0,
\end{equation}
an integrable system introduced in \cite{MX}, where it was proved to satisfy the second order, but not the first order, integrability conditions, to posses a $3\times 3$ Lax pair and to be a degeneration of the discrete integrable Tzitzeica equation proposed by Adler in \cite{A}. Moreover this equation has the singularity confinement property, can be trilinearized and possesses multisoliton solutions, \cite{GRSS}.

Finally, if in (\ref{Quadrelli1}), when $\tau\not=0$, $1$ we apply the (not allowed) transformation $v_{n,m}\coloneqq \frac{1-\tau}{\tau}w_{n,m}-1$, we obtain
\begin{equation}
\label{jk3}w_{n,m}w_{n+1,m}+w_{n,m+1}w_{n+1,m+1}+w_{n+1,m}w_{n,m+1}\left(1+w_{n,m}+w_{n+1,m+1}\right)+\chi=0, 
\end{equation}
where $\chi\coloneqq\frac{\left(\tau-3\right)\tau^2}{\left(1-\tau\right)^3}$, which doesn't satisfy the first order integrability conditions for three-point generalized symmetries either autonomous or not, either in direction $n$ or $m$. In \cite{SHL} we showed the integrability of the subcase $\chi=0$ constructing two five-point generalized symmetries, one in the $n$ direction and the other one in the $m$ direction, and a $3\times 3$ Lax pair. An indication of the integrability of the general case (\ref{jk3}) for arbitrary $\chi$ was provided showing its algebraic entropy vanishes. Other strong indications of integrability for arbitrary $\chi$, such as the singularity confinement property, bilinear form, multisoliton solutions and a continuous limit into the \emph{mKdV} equation when $\chi=-1$, were established in \cite{GRSS}. In the case $\chi=-1$ we can provide the following five-point symmetry in the $n$ direction depending on the points $\left(n+2,m\right)$, $\left(n+1,m\right)$, $\left(n,m\right)$, $\left(n-1,m\right)$ and $\left(n-2,m\right)$:
\[
w_{n,m,t}=\frac{w_{n,m}\left(w_{n,m}+1\right)\left(w_{n,m}w_{n-1,m}-1\right)\left(w_{n+1,m}w_{n,m}-1\right)\left(w_{n+2,m}w_{n+1,m}-w_{n-1,m}w_{n-2,m}\right)}{\left(w_{n,m}w_{n-1,m}w_{n-2,m}+1\right)\left(w_{n+1,m}w_{n,m}w_{n-1,m}+1\right)\left(w_{n+1,m}w_{n+1,m}w_{n,m}+1\right)},
\]
where $t$ is a group parameter. The last generalized symmetry is invariant under $\tilde w_{n,m}\coloneqq 1/w_{n,m}$ and under the following Miura transformation
\bea
z_{n,m}\coloneqq\frac{w_{n+1,m}w_{n,m}-1}{w_{n+1,m}w_{n,m}w_{n-1,m}+1}\label{ivecloga}
\eea
it can be transformed into a Bogoyavlenskyi lattice
\bea
\nonumber z_{n,m,t}=z_{n,m}\left(z_{n,m}+1\right)\left(z_{n+2,m}z_{n+1,m}-z_{n-1,m}z_{n-2,m}\right).
\eea
\begin{prop}
If (\ref{jk3}) with $\chi=-1$ is satisfied, given (\ref{ivecloga}), then
\begin{subequations}\label{Sibilla}
\begin{gather}
w_{n+1,m}=-\frac{z_{n,m}+1}{\left(z_{n,m}w_{n-1,m}-1\right)w_{n,m}},\label{Sibilla1}\\
w_{n,m+1}=-\frac{\left(z_{n,m+1}+z_{n,m}+1\right)\left(w_{n-1,m}+1\right)w_{n,m}}{\left(z_{n,m}+1\right)\left(w_{n,m}+1\right)},\label{Sibilla2}\\
w_{n-1,m+1}=\frac{\left(z_{n,m}+1\right)\left(w_{n,m}w_{n-1,m}-1\right)}{z_{n,m+1}\left(w_{n-1,m}+1\right)w_{n,m}},\label{Sibilla3}
\end{gather}
\end{subequations}
and $z_{n,m}$ satisfies
\bea
z_{n,m}\left(z_{n+1,m}+1\right)+z_{n+1,m+1}\left(z_{n,m+1}+z_{n,m}+1\right)=0.\label{Sibilla4}
\eea
\end{prop}
To prove (\ref{Sibilla1}), just solve (\ref{ivecloga}) with respect to $w_{n+1,m}$; to obtain (\ref{Sibilla2}), substitute (\ref{Sibilla1}) and its shifted once along direction $m$ into the equation (\ref{jk3}) with $\chi=-1$, solve with respect to $w_{n-1,m+1}$, substitute this result into the equation (\ref{jk3}) with $\chi=-1$ shifted back once along direction $n$ and solve with respect to $w_{n,m+1}$; (\ref{Sibilla3}) follows inserting (\ref{Sibilla2}) into the previous result for $w_{n-1,m+1}$. The three relations (\ref{Sibilla}) provide a Miura transformation between equation (\ref{jk3}) with $\chi=-1$ and (\ref{Sibilla4}): the compatibility between the $z-$variables implies (\ref{jk3}) with $\chi=-1$, while the compatibility between the $w-$variables implies (\ref{Sibilla4}).\\
\indent Equation (\ref{Sibilla4}) is an integrable lattice possessing a $3\times 3$ Lax representation, \cite{SHL}. When $z_{n,m}\not=0$, under the inversion $\tilde w_{n,m}\coloneqq 1/z_{n,m}$ the equation (\ref{Sibilla4}) is mapped into the equation (\ref{jk3}) with $\chi=0$, so (\ref{ivecloga}) provides also a Miura mapping from $w_{n,m}$ solving (\ref{jk3}) with $\chi=-1$ to $\tilde w_{n,m}$ solving the same equation but with $\chi=0$. This Miura transformation induces, through the mapping $v_{n,m}\coloneqq \frac{1-\tau}{\tau}w_{n,m}-1$, a corresponding Miura transformation from a solution $v_{n,m}$ of (\ref{Quadrelli1}) with $\tau=1/3$ to a solution $\tilde v_{n,m}$ of (\ref{Quadrelli1}) with $\tau=3$. Another set of Miura transformations between the equations (\ref{kj1}), (\ref{jk2}) and (\ref{jk3}) was derived in \cite{GRSS}.\\
\indent Summing up, we have very strong indications of integrability for the master equation (\ref{Quadrelli1}) which, when $\tau=3$, $1/3$, has a continuous limit into the \emph{mKdV} equation, \cite{GRSS}.\\
\indent If in (\ref{Quadrelli2}), (\ref{Quadrelli3}) we apply the (not allowed) transformations $w_{n,m}\coloneqq\delta sgn\left(\epsilon\right)/v_{n,m}$ and $\tilde w_{n,m}\coloneqq\delta/v_{n,m}$ respectively, we obtain
\begin{subequations}
\bea
\label{k1}&&\frac{\tau}{\left|\epsilon\right|}+w_{n,m}+w_{n+1,m+1}+\frac{1}{\epsilon}\left(w_{n+1,m}+w_{n,m+1}\right)\\
\nonumber && \quad+\delta\left[\frac{1}{\epsilon}w_{n+1,m}w_{n,m+1}\left(w_{n,m}+w_{n+1,m+1}\right)+w_{n,m}w_{n+1,m+1}\left(w_{n+1,m}+w_{n,m+1}\right)\right]=0,\\
\label{k2}&&\tau+\tilde w_{n,m}+\tilde w_{n+1,m+1}+\epsilon\left(\tilde w_{n+1,m}+\tilde w_{n,m+1}\right)\\
\nonumber &&\quad+\delta\left[\tilde w_{n+1,m}\tilde w_{n,m+1}\left(\tilde w_{n,m}+\tilde w_{n+1,m+1}\right)+\epsilon \tilde w_{n,m}\tilde w_{n+1,m+1}\left(\tilde w_{n+1,m}+\tilde w_{n,m+1}\right)\right]=0.
\eea
\end{subequations}
Eqs.~(\ref{k1}, \ref{k2}) are just an almost trivial looking  modification of the two integrable systems discussed in \cite{SGR}, which are recovered when $\tau=0$. In that paper it was shown that, when $\tau=0$, Eqs.~(\ref{k2}, \ref{k1}) are mapped through a M\"obious transformation respectively to the Hirota discrete sine-Gordon equation and to its potential form. After in (\ref{k1}) we replace $\epsilon\rightarrow 1/\epsilon$ and in (\ref{k2}) $\delta\rightarrow s\delta$, with $s\coloneqq sgn\left(\epsilon\right)$, the precise form of the potentiation induced between them is
\bea
\nonumber w_{n,m}=|\epsilon|^{1/2}\frac{\tilde w_{n+1,m}+\tilde w_{n,m+1}}{1+s\delta\tilde w_{n+1,m}\tilde w_{n,m+1}}.
\eea 
These equations satisfy the first order integrability conditions for three-point generalized symmetries either autonomous or not if and only if $\tau=0$, which in this limit, in the $n$ direction, are respectively given by
\begin{gather*}
    w_{n,m,t}=\frac{\left(\delta w_{n,m}^2-\epsilon\right)\left(\delta\epsilon w_{n,m}^2-1\right)\left(w_{n+1,m}-w_{n-1,m}\right)}{\left(1+\delta w_{n,m}w_{n+1,m}\right)\left(1+\delta w_{n,m}w_{n-1,m}\right)},
    \\
    \tilde w_{n,m,\tilde t}=Y\frac{\left(\delta\tilde w_{n,m}^2-1\right)\left(\tilde w_{n+1,m}-\tilde w_{n-1,m}\right)}{\delta\tilde w_{n+1,m}\tilde w_{n-1,m}-1}+\left[\left(-1\right)^n\kappa+\left(-1\right)^m\theta\right]\left(\delta\tilde w_{n,m}^2-1\right),
\end{gather*}
where $t$ and $\tilde t$ are two group parameters, and, in the $m$ direction, by similar expressions obtained changing $w_{n+1,m}\rightarrow w_{n,m+1}$ and $w_{n-1,m}\rightarrow w_{n,m-1}$. The second integrable system shows a two parameters non autonomous point symmetry tail too. We note that both the integrable systems are invariant under $w_{n,m}\coloneqq -v_{n,m}$; the first integrable system is covariant under the inversion $w_{n,m}\coloneqq 1/v_{n,m}$ as $\epsilon$ is changed into $1/\epsilon$, while the second one is invariant; under the non autonomous transformation $w_{n,m}\coloneqq \left(-1\right)^{n+m}v_{n,m}$ both the integrable systems are covariant as in the first case $\epsilon$ is changed into $-\epsilon$ and $\delta$ into $-\delta$, while in the second one $\epsilon$ is changed into $-\epsilon$. This implies that in those systems we can limit ourselves to the range $0<\epsilon<1$. Moreover the second integrable system under the non autonomous transformation $\tilde w_{n,m}\coloneqq (v_{n,m})^{\left(-1\right)^{n+m}}$ is invariant when $\delta=1$ and covariant when $\delta=-1$ as $\epsilon$ is changed into $\delta\epsilon$. Finally both the integrable systems are covariant under the transformation $w_{n,m}\coloneqq\ri v_{n,m}$ as $\delta$ is changed into $-\delta$. This implies that in those systems we can always take $\delta=1$ but in general, if we allow such a transformation, the solution will be no more a real field but a complex one. Let's also note that the non autonomous transformation $w_{n,m}\coloneqq \left(-1\right)^{n}v_{n,m}$ or $w_{n,m}\coloneqq \left(-1\right)^{m}v_{n,m}$ brings both the integrable systems from class $\CQ^+$ into class $\CQ^-$.

An indication that the general cases (\ref{k1}, \ref{k2}) are not integrable when $\tau\not=0$ can be obtained showing their algebraic entropy~\cite{AE} doesn't vanish.

If in (\ref{Quadrelli4}) we apply the (not allowed) transformation $v_{n,m}\coloneqq\frac{|\rho|^{1/2}w_{0,0}+1}{|\rho|^{1/2}w_{0,0}-1/\epsilon}$, with $\rho\coloneqq\frac{-1+2\epsilon}{\epsilon\left(\epsilon-2\right)}\not=0$, we obtain
\begin{multline}
\label{jk4}
w_{n,m}+w_{n+1,m+1}+w_{n,m}w_{n+1,m+1}\left[\delta\left(w_{n+1,m}+w_{n,m+1}\right)+\epsilon|\rho|^{3/2}w_{n+1,m}w_{n,m+1}\right]
\\
{}+\frac{\delta}{\epsilon|\rho|^{3/2}}=0,
\end{multline}
where $\delta\coloneqq sgn\left(\rho\right)=sgn\left(1/\epsilon-2\right)$, which, for $\delta=-1$, is a real discrete Tzitzeica equation with coefficient $c=1/\left(\epsilon|\rho|^{3/2}\right)$ and for $\delta=1$, through the (not allowed) transformation $w_{n,m}\rightarrow\ri w_{n,m}$ becomes a complex Tzitzeica equation with coefficient $c=\ri/\left(\epsilon|\rho|^{3/2}\right)$. We remember that the Tzitzeica equation possesses a $3\times 3$ Lax representation and satisfies \cite{MX} the second order, but not the first order, integrability conditions.

We note that, besides not being subcases of the $Q_{V}$ equation, our systems (\ref{Quadrelli}), except for (\ref{Quadrelli2}, \ref{Quadrelli3}) with $\tau=0$, where a five-point generalized symmetry depending on the points $\left(n+1,m\right)$, $\left(n,m+1\right)$, $\left(n,m\right)$, $\left(n-1,m\right)$ and $\left(n,m-1\right)$ exists, are not included into the Garifullin-Yamilov class \cite{GY} too.

\section{Concluding remarks}\label{s3}
In this paper we have considered the application of a multiple scale expansion to a class of dispersive multilinear partial difference equation on the square lattice, $\CQ^+$.  The integrability conditions we obtain when we require that the multiple scale expansion of the discrete class of equations is equivalent to the equations of the \emph{NLS} hierarchy reduce the $\CN=13$ initial parameters defining the $\CQ^+$ class to a maximum of $\CN=2$ free (continuous) parameters defining four equations. A great effort has been directed to extend the expansion up to order $\ep^6$, the related integrability conditions appearing in this paper for the first time. As a result of our efforts we have been able to compare the $A_3$ integrable equations to the $A_4$ integrable equations. They turn out to be the same, so that one could presume we might be already in the asymptotic regime and that the obtained equations might be integrable. However a non vanishing algebraic entropy is an indication that the general cases (\ref{k1}, \ref{k2}) are not integrable when $\tau\not=0$.

An open problem seems of major importance now: the consideration of the second class of dispersive multilinear partial difference equations on the square lattice, $\CQ^-$ is a major task which will provide by sure new classes of integrable equations. However in this case the lowest order integrability conditions appear already at order $\ep^2$ and will not be an equation of the \emph{NLS} type but more likely a coupled wave equations. Work is in progress on it.

\section*{Acknowledgments}
LD and CS were partly supported by the Italian Ministry of Education and Research, PRIN
``Nonlinear waves: integrable fine dimensional reductions and discretizations" from 2007
to 2009 and PRIN ``Continuous and discrete nonlinear integrable evolutions: from water
waves to symplectic maps" from 2010. RHH is supported by projects PID2019-106802GB-I00 and 
PID2021-124473NB-I00 from the Ministry of Science and Innovation of Spain, and would like to thank the INFN, Sezione Roma Tre and the UPM for their support during his visits to Rome. We thank Matteo Petrera for many enlightening discussions in the first stages of this paper.

Before the publication of this work, Professor Decio Levi untimely passed away. Christian Scimiterna and Rafael Hernandez Heredero would like to express their deep admiration to their Professor, to whom they owe so many things in their personal and academic lifes, and whom they miss with deep emotion.

\section*{Appendix}
\allowdisplaybreaks
Here we present explicitly the $48$ conditions for $\varepsilon^6$ $S$-asymptotic integrability ($A_{4}$-integrability) involving the real ($S_{j}$) and imaginary parts ($T_{j}$) of the coefficients $\eta_{j}$, $j=1$,\ldots, $34$ of the differential polynomial (\ref{f24}). The expressions of the coefficients $\kappa_{m}$ , $m=1$,\ldots, $77$ of the differential polynomial (\ref{f34}) as functions of the $\eta_{j}$, $j=1$,\ldots, $34$ are complicated, so we will omit them.

{\tiny\bea\label{Phoenix}
&&T_{2}=\left(\frac{a}{11}+\frac{3b}{4}\right)\frac{S_{10}}{\rho_{2}}+\left(\frac{13a}{11}+\frac{b}{2}\right)\frac{S_{18}-S_{15}}{2\rho_{1}}+\left(\frac{37a}{11}+b\right)\frac{I_{6}S_{27}}{8\rho_{1}\rho_{2}}+T_{1}+\left(\frac{a^2}{2}+\frac{13ab}{11}+\frac{b^2}{2}\right)\frac{T_{22}}{4\rho_{1}\rho_{2}}\\
&&+\left[I_{1}+\left(37a^2-15ab-11b^2\right)\frac{I_{8}}{44\rho_{1}\rho_{2}}\right]\frac{T_{27}}{2\rho_{2}}+\left[\left(\frac{13a}{11}+\frac{b}{2}\right)\frac{R_{2}}{8}-\left(32a^2+27ab-189b^2\right)\frac{I_{6}}{352}+\left(\frac{a}{11}+\frac{b}{4}\right)\frac{aI_{12}}{8}\right]\frac{T_{32}}{\rho_{1}\rho_{2}}-\nonumber\\
&&-\left[\left(9a-\frac{7b}{2}\right)\frac{R_{12}}{22\rho_{1}\rho_{2}}+\frac{I_{1}}{\rho_{2}}+\frac{7I_{2}-29I_{3}-14I_{5}}{44\rho_{1}}+\left(39a^2-53ab+\frac{67b^2}{2}\right)\frac{I_{8}}{44\rho_{1}\rho_{2}^2}\right]\frac{T_{33}}{2},\nonumber\\
&&T_{3}=\frac{a}{2\rho_{2}}\left(\frac{\rho_{1}}{\rho_{2}}S_{10}+S_{15}-S_{18}+S_{20}\right)-\left(a-\frac{b}{2}\right)\frac{I_{6}S_{27}}{2\rho_{2}^2}-\frac{abT_{22}}{4\rho_{2}^2}+\left[\frac{aR_{12}}{2\rho_{2}}-I_{3}-\left(a-b\right)\frac{aI_{8}}{2\rho_{2}^2}\right]\frac{T_{27}}{2\rho_{2}}-\nonumber\\
&&-\left[R_{2}+\frac{\left(3a+b\right)I_{6}+aI_{12}}{2\rho_{2}}\right]\frac{aT_{32}}{8\rho_{2}^2}-\left[\left(3a+b\right)\frac{R_{12}}{4\rho_{2}}+\frac{\rho_{1}I_{1}}{\rho_{2}}-\frac{I_{2}+3I_{3}+2I_{5}}{4}-\left(a^2-3ab+2b^2\right)\frac{I_{8}}{4\rho_{2}^2}\right]\frac{T_{33}}{2\rho_{2}},\nonumber\\
&&T_{4}=\left(\frac{35a}{33}+\frac{b}{2}\right)\frac{\rho_{1}S_{10}}{2\rho_{2}^2}-\left(\frac{35a}{33}-\frac{b}{2}\right)\frac{S_{15}}{2\rho_{2}}+\left(\frac{34a}{33}+\frac{b}{4}\right)\frac{S_{18}}{\rho_{2}}-\frac{aS_{20}}{2\rho_{2}}+\left[\frac{R_{2}}{2}+\frac{67aI_{6}}{33\rho_{2}}+\left(\frac{a}{2}+b\right)\frac{I_{12}}{2\rho_{2}}\right]\frac{S_{27}}{2\rho_{2}}+\frac{\rho_{1}T_{1}}{\rho_{2}}+\nonumber\\
&&+\left(\frac{a^2}{2}+\frac{101ab}{33}+\frac{b^2}{2}\right)\frac{T_{22}}{4\rho_{2}^2}+\left[-\left(a-b\right)\frac{R_{12}}{8\rho_{2}}+\frac{\rho_{1}I_{1}}{\rho_{2}}-\frac{I_{2}-I_{3}-2I_{5}}{8}+\left(\frac{17a^2}{33}-\frac{103ab}{264}+\frac{b^2}{8}\right)\frac{I_{8}}{\rho_{2}^2}\right]\frac{T_{27}}{\rho_{2}}+\nonumber\\
&&+\left[\left(\frac{101a}{33}+\frac{3b}{2}\right)\frac{R_{2}}{2}-\left(\frac{19a^2}{3}+\frac{97ab}{4}-\frac{591b^2}{4}\right)\frac{I_{6}}{22\rho_{2}}+\left(\frac{17a^2}{33}+\frac{3ab}{8}+\frac{b^2}{2}\right)\frac{I_{12}}{\rho_{2}}\right]\frac{T_{32}}{4\rho_{2}^2}+\left[\left(\frac{37a}{2}-35b\right)\frac{R_{12}}{66\rho_{2}}+\right.\nonumber\\
&&\left.+\frac{35I_{2}-2I_{3}-70I_{5}}{66}-\left(\frac{25a^2}{3}-\frac{317ab}{6}+\frac{323b^2}{4}\right)\frac{I_{8}}{22\rho_{2}}\right]\frac{T_{33}}{2\rho_{2}},\ \ \ S_{5}=-\frac{\rho_{1}}{\rho_{2}}\left(S_{1}+\frac{3S_{2}}{2}\right)+\frac{9}{2}\left(S_{3}+S_{4}\right)-S_{6}+\frac{S_{7}}{2}-\nonumber\\
&&-\left(a^2+\frac{7ab}{2}+\frac{b^2}{4}\right)\frac{S_{22}}{2\rho_{2}^2}+\left[-\left(5a-7b\right)\frac{R_{12}}{16\rho_{2}}-\frac{7\rho_{1}I_{1}}{4\rho_{2}}-\frac{7I_{2}-35I_{3}+18I_{5}}{16}-\left(\frac{7a}{16}+b\right)\frac{aI_{8}}{\rho_{2}^2}\right]\frac{S_{27}}{\rho_{2}}+\left(a+7b\right)\frac{\rho_{1}T_{10}}{4\rho_{2}^2}+\nonumber\\
&&+\left(a-\frac{5b}{2}\right)\frac{T_{15}}{2\rho_{2}}+\left(a+\frac{3b}{2}\right)\frac{T_{18}}{2\rho_{2}}+\left(a+b\right)\frac{T_{20}}{4\rho_{2}}-\left[3R_{2}+\left(a-23b\right)\frac{I_{6}}{2\rho_{2}}+\left(\frac{a}{2}+b\right)\frac{3I_{12}}{\rho_{2}}\right]\frac{T_{27}}{8\rho_{2}}+\nonumber\\
&&+\left\{\left(\frac{a^2}{2}-ab+3b^2\right)\frac{R_{12}}{8\rho_{2}}+\frac{\rho_{1}aI_{1}}{\rho_{2}}+\frac{\left(3a-5b\right)I_{2}+\left(a+39b\right)I_{3}}{16}+\left(\frac{a}{4}-b\right)\frac{I_{5}}{2}+\left[\frac{a\left(5a+13b\right)}{2}-11b^2\right]\frac{aI_{8}}{8\rho_{2}^2}\right\}\frac{T_{32}}{\rho_{2}^2}-\nonumber\\
&&-\left[7R_{2}+\left(7a+15b\right)\frac{I_{6}}{2\rho_{2}}+\left(\frac{a}{2}+b\right)\frac{7I_{12}}{\rho_{2}}\right]\frac{T_{33}}{8\rho_{2}},\ \ \ T_{5}=-\left(\frac{31a}{11}-3b\right)\frac{\rho_{1}S_{10}}{4\rho_{2}^2}+\left(\frac{31a}{11}-5b\right)\frac{S_{15}}{4\rho_{2}}-\left(\frac{53a}{11}-3b\right)\frac{S_{18}}{4\rho_{2}}+\nonumber\\
&&+\left(a-\frac{b}{2}\right)\frac{S_{20}}{\rho_{2}}-\left[\frac{R_{2}}{2}+\left(\frac{97a}{44}-b\right)\frac{I_{6}}{\rho_{2}}+\left(\frac{a}{2}+b\right)\frac{I_{12}}{2\rho_{2}}\right]\frac{S_{27}}{2\rho_{2}}-\frac{\rho_{1}T_{1}}{\rho_{2}}+T_{6}-\left(3a^2+\frac{31ab}{11}-b^2\right)\frac{T_{22}}{8\rho_{2}^2}+\nonumber\\
&&+\left\{\left(a-\frac{b}{2}\right)\frac{R_{12}}{2\rho_{2}}-\frac{\rho_{1}I_{1}}{2\rho_{2}}+\frac{I_{2}-I_{3}-4I_{5}}{4}-\left[\frac{a\left(8a-19b\right)}{11}+\frac{9b^2}{8}\right]\frac{I_{8}}{\rho_{2}^2}\right\}\frac{T_{27}}{\rho_{2}}-\left[\left(\frac{75a}{11}-b\right)R_{2}+\left(\frac{49a^2}{2}-205ab+\right.\right.\nonumber\\
&&\left.\left.+\frac{615b^2}{2}\right)\frac{I_{6}}{11\rho_{2}}+\left(\frac{9a}{11}+7b\right)\frac{aI_{12}}{2\rho_{2}}\right]\frac{T_{32}}{16\rho_{2}^2}-\left\{\left(\frac{9a}{2}+b\right)\frac{R_{12}}{44\rho_{2}}+\frac{\rho_{1}I_{1}}{2\rho_{2}}-\frac{I_{2}-12I_{3}+20I_{5}}{44}-\left[13a^2-\frac{\left(585a-533b\right)b}{8}\right]\frac{I_{8}}{22\rho_{2}^2}\right\}\frac{T_{33}}{\rho_{2}},\nonumber
\eea
\bea
&&T_{7}=-\left(\frac{23a}{11}+\frac{b}{2}\right)\frac{\rho_{1}S_{10}}{\rho_{2}^2}+\left(\frac{23a}{11}-\frac{b}{2}\right)\frac{S_{15}}{\rho_{2}}-\left(\frac{34a}{11}+\frac{b}{2}\right)\frac{S_{18}}{\rho_{2}}+\frac{3aS_{20}}{2\rho_{2}}-\left[R_{2}+\left(\frac{123a}{11}-b\right)\frac{I_{6}}{2\rho_{2}}+\left(\frac{a}{2}+b\right)\frac{I_{12}}{\rho_{2}}\right]\frac{S_{27}}{2\rho_{2}}-\nonumber\\
&&-\frac{2\rho_{1}T_{1}}{\rho_{2}}+T_{6}-\left(\frac{a^2}{2}+\frac{45ab}{11}+\frac{b^2}{2}\right)\frac{T_{22}}{2\rho_{2}^2}+\left[\left(a-\frac{b}{2}\right)\frac{R_{12}}{2\rho_{2}}-\frac{2\rho_{1}I_{1}}{\rho_{2}}+\frac{I_{2}-3I_{3}-2I_{5}}{4}-\left(\frac{17a^2}{11}-\frac{57ab}{44}+\frac{b^2}{4}\right)\frac{I_{8}}{\rho_{2}^2}\right]\frac{T_{27}}{\rho_{2}}-\nonumber\\
&&-\left\{3\left(\frac{15a}{11}+\frac{b}{2}\right)R_{2}+\left[47a^2-\frac{\left(107a-815b\right)b}{2}\right]\frac{I_{6}}{22\rho_{2}}+\left(\frac{17a^2}{11}+\frac{3ab}{4}+b^2\right)\frac{I_{12}}{\rho_{2}}\right\}\frac{T_{32}}{4\rho_{2}^2}-\left[\left(81a-59b\right)\frac{R_{12}}{44\rho_{2}}+\frac{\rho_{1}I_{1}}{\rho_{2}}+\right.\nonumber\\
&&\left.+\frac{59\left(I_{2}-I_{3}\right)-162I_{5}}{44}-\left(\frac{105a^2}{4}-71ab+\frac{457b^2}{4}\right)\frac{I_{8}}{11\rho_{2}^2}\right]\frac{T_{33}}{2\rho_{2}},\ \ \ S_{8}=-\frac{2\rho_{1}S_{2}}{\rho_{2}}+3S_{3}+6S_{4}-2S_{6}+S_{7}-\nonumber\\
&&-\left(\frac{a^2}{2}+ab+\frac{b^2}{2}\right)\frac{S_{22}}{\rho_{2}^2}+\left[\left(a+b\right)\frac{R_{12}}{4\rho_{2}}-\frac{\rho_{1}I_{1}}{\rho_{2}}-\frac{I_{2}-5I_{3}+6I_{5}}{4}-\left(\frac{a}{4}+b\right)\frac{aI_{8}}{\rho_{2}^2}\right]\frac{S_{27}}{\rho_{2}}+\left(a+b\right)\frac{\rho_{1}T_{10}}{\rho_{2}^2}-\frac{b}{\rho_{2}}\left(T_{15}-T_{18}\right)-\nonumber\\
&&-\left[R_{2}+\left(a-5b\right)\frac{I_{6}}{2\rho_{2}}+\left(\frac{a}{2}+b\right)\frac{I_{12}}{\rho_{2}}\right]\frac{T_{27}}{2\rho_{2}}+\left[a\left(a+b\right)\frac{R_{12}}{4\rho_{2}}+\frac{\rho_{1}aI_{1}}{\rho_{2}}+\frac{b\left(I_{2}+5I_{3}-4I_{5}\right)}{4}+a\left(\frac{a^2}{2}+\frac{ab}{4}-b^2\right)\frac{I_{8}}{\rho_{2}^2}\right]\frac{T_{32}}{\rho_{2}^2}-\nonumber\\
&&-\left[R_{2}+\left(a+3b\right)\frac{I_{6}}{2\rho_{2}}+\left(\frac{a}{2}+b\right)\frac{I_{12}}{\rho_{2}}\right]\frac{T_{33}}{2\rho_{2}},\ \ \ T_{8}=-\left(\frac{31a}{11}-b\right)\frac{\rho_{1}S_{10}}{2\rho_{2}^2}+\left(\frac{31a}{11}-3b\right)\frac{S_{15}}{2\rho_{2}}-\left(\frac{53a}{11}-b\right)\frac{S_{18}}{2\rho_{2}}+\nonumber\\
&&+\left(2a-b\right)\frac{S_{20}}{\rho_{2}}-\left[R_{2}+\left(\frac{97a}{22}-b\right)\frac{I_{6}}{\rho_{2}}+\left(\frac{a}{2}+b\right)\frac{I_{12}}{\rho_{2}}\right]\frac{S_{27}}{2\rho_{2}}-\frac{2\rho_{1}T_{1}}{\rho_{2}}+T_{6}-\left(a^2+\frac{75ab}{44}-b^2\right)\frac{T_{22}}{4\rho_{2}^2}+\nonumber\\
&&+\left[\left(a-b\right)\frac{3R_{12}}{4\rho_{2}}-\frac{2\rho_{1}I_{1}}{\rho_{2}}+\frac{3\left(I_{2}-I_{3}-2I_{5}\right)}{4}-\left(\frac{53a^2}{44}-\frac{16ab}{11}+\frac{3b^2}{4}\right)\frac{I_{8}}{\rho_{2}^2}\right]\frac{T_{27}}{\rho_{2}}-\left[\left(\frac{75a}{11}+b\right)\frac{R_{2}}{8}-\right.\nonumber\\
&&\left.-\left(\frac{39a^2}{16}+5ab-\frac{461b^2}{16}\right)\frac{I_{6}}{11\rho_{2}}+\left(\frac{53a^2}{44}+\frac{ab}{4}+b^2\right)\frac{I_{12}}{4\rho_{2}}\right]\frac{T_{32}}{\rho_{2}^2}-\left\{\left(5a-\frac{31b}{4}\right)\frac{R_{12}}{11\rho_{2}}+\frac{31I_{2}-9I_{3}-62I_{5}}{44}-\right.\nonumber\\
&&\left.-\left[19a^2-\frac{\left(255a-313b\right)b}{2}\right]\frac{I_{8}}{44\rho_{2}^2}\right\}\frac{T_{33}}{\rho_{2}},\nonumber\\
&&S_{9}=\frac{S_{10}}{2}-\frac{\rho_{2}}{2\rho_{1}}\left(S_{15}-S_{18}\right)+\frac{I_{6}S_{27}+bT_{22}}{4\rho_{1}}+\left(a-b\right)\frac{I_{8}T_{27}}{4\rho_{1}\rho_{2}}+\left[R_{2}+\frac{\left(3a-7b\right)I_{6}+aI_{12}}{2\rho_{2}}\right]\frac{T_{32}}{8\rho_{1}}+\frac{3bI_{8}T_{33}}{4\rho_{1}\rho_{2}},\nonumber\\
&&T_{9}=T_{10}-\frac{I_{1}T_{32}}{2\rho_{2}}-\frac{I_{6}T_{33}}{4\rho_{1}},\ \ \ S_{11}=\frac{S_{22}}{2}-\left[\frac{R_{12}}{2}+\left(a-b\right)\frac{I_{8}}{\rho_{2}}\right]\frac{T_{32}}{2\rho_{2}},\ \ \ T_{11}=T_{22}+\frac{I_{8}T_{27}-\left(I_{6}+I_{12}\right)T_{32}}{2\rho_{2}},\nonumber\\
&&S_{12}=-\frac{I_{8}S_{27}}{\rho_{2}}+\left(\frac{a}{2}-b\right)\frac{I_{8}T_{32}}{\rho_{2}^2},\ \ \ T_{12}=-\frac{I_{8}\left(T_{27}-T_{33}\right)}{\rho_{2}},\ \ \ S_{13}=-\frac{2\rho_{1}S_{10}}{\rho_{2}}+2S_{15}-S_{18}+S_{20}-\frac{I_{6}S_{27}}{\rho_{2}}-\nonumber\\
&&-\frac{aT_{22}}{\rho_{2}}-2\left(a-b\right)\frac{I_{8}T_{27}}{\rho_{2}^2}-\left[R_{2}+\left(5a-7b\right)\frac{I_{6}}{2\rho_{2}}-\left(\frac{a}{2}-b\right)\frac{I_{12}}{\rho_{2}}\right]\frac{T_{32}}{2\rho_{2}}+\left(3a-4b\right)\frac{I_{8}T_{33}}{\rho_{2}^2},\ \ \ T_{13}=T_{18}+T_{20}+\nonumber\\
&&+\left[\left(a+b\right)\frac{R_{12}}{4\rho_{2}}+\frac{\rho_{1}I_{1}}{\rho_{2}}-\frac{I_{2}-I_{3}+2I_{5}}{4}+\left(\frac{a^2}{2}+ab-b^2\right)\frac{I_{8}}{2\rho_{2}^2}\right]\frac{T_{32}}{\rho_{2}},\ \ \ S_{14}=0,\ \ \ T_{14}=-\frac{I_{8}T_{32}}{2\rho_{2}},\nonumber\\
&&S_{16}=-\frac{I_{6}S_{27}+aT_{22}}{2\rho_{2}}+\left(a-\frac{3b}{2}\right)\frac{I_{6}T_{32}}{2\rho_{2}^2}+\frac{R_{12}T_{33}}{2\rho_{2}},\ \ \ T_{16}=\frac{aS_{22}-I_{6}T_{27}}{2\rho_{2}}+\left(\frac{aR_{12}}{2\rho_{2}}-I_{3}\right)\frac{T_{32}}{2\rho_{2}}+\left(I_{6}+I_{12}\right)\frac{T_{33}}{2\rho_{2}},\nonumber\\
&&S_{17}=S_{15}+\frac{I_{12}S_{27}}{2\rho_{2}}-\left(a-b\right)\frac{I_{8}T_{27}}{2\rho_{2}^2}+\left[\left(2a+3b\right)I_{6}+bI_{12}\right]\frac{T_{32}}{4\rho_{2}^2}+\left(a-\frac{5b}{4}\right)\frac{I_{8}T_{33}}{\rho_{2}^2},\ \ \ T_{17}=-\frac{aS_{22}}{2\rho_{2}}+\nonumber\\
&&+\left(R_{12}-\frac{bI_{8}}{\rho_{2}}\right)\frac{S_{27}}{2\rho_{2}}+\frac{\rho_{1}T_{10}}{\rho_{2}}+\left[\left(a+3b\right)\frac{R_{12}}{4\rho_{2}}+\frac{\rho_{1}I_{1}}{\rho_{2}}-\frac{I_{2}-I_{3}-2I_{5}}{4}+\left(\frac{3a^2}{2}-ab-b^2\right)\frac{I_{8}}{2\rho_{2}^2}\right]\frac{T_{32}}{2\rho_{2}},\ \ \ S_{19}=S_{20},\nonumber\\
&&T_{19}=T_{20},\ \ \ S_{21}=S_{22},\ \ \ T_{21}=T_{22},\ \ \ S_{23}=-\frac{2\rho_{1}S_{10}}{\rho_{2}}+2S_{15}-S_{18}+S_{20}-\left(I_{6}-\frac{I_{12}}{2}\right)\frac{S_{27}}{\rho_{2}}-\left(a+b\right)\frac{T_{22}}{2\rho_{2}}+\nonumber\\
&&+\left[R_{12}-\left(a-b\right)\frac{I_{8}}{\rho_{2}}\right]\frac{T_{27}}{2\rho_{2}}-\left[R_{2}-\left(\frac{a}{2}-3b\right)\frac{I_{6}}{\rho_{2}}+\left(a-b\right)\frac{I_{12}}{2\rho_{2}}\right]\frac{T_{32}}{2\rho_{2}}-\left(R_{12}-\frac{3bI_{8}}{2\rho_{2}}\right)\frac{T_{33}}{2\rho_{2}},\nonumber\\
&&T_{23}=\left(a-\frac{b}{2}\right)\frac{S_{22}}{\rho_{2}}-\left(R_{12}-\frac{bI_{8}}{\rho_{2}}\right)\frac{S_{27}}{2\rho_{2}}-\frac{\rho_{1}T_{10}}{\rho_{2}}+T_{15}+T_{18}+\frac{I_{12}T_{27}}{2\rho_{2}}+\left[\left(\frac{a}{2}-b\right)\frac{R_{12}}{\rho_{2}}+\frac{I_{2}-I_{3}-4I_{5}}{2}\right]\frac{T_{32}}{2\rho_{2}}-\nonumber\\
&&-\left(\frac{a}{2}-b\right)\frac{aI_{8}T_{32}}{2\rho_{2}^3}-\frac{I_{12}T_{33}}{2\rho_{2}},\ \ \ S_{24}=S_{18}+S_{20}-\frac{\left(I_{6}-I_{12}\right)S_{27}+bT_{22}}{2\rho_{2}}+\left[R_{12}+\left(a-b\right)\frac{I_{8}}{\rho_{2}}\right]\frac{T_{27}}{2\rho_{2}}+\nonumber\\
&&+\left[\left(a-\frac{3b}{2}\right)I_{6}+\frac{bI_{12}}{4}\right]\frac{T_{32}}{\rho_{2}^2}-\left(a-\frac{5b}{4}\right)\frac{I_{8}T_{33}}{\rho_{2}^2},\ \ \ T_{24}=\left(a+b\right)\frac{S_{22}}{2\rho_{2}}+\left(R_{12}+\frac{bI_{8}}{\rho_{2}}\right)\frac{S_{27}}{2\rho_{2}}-\frac{\rho_{1}T_{10}}{\rho_{2}}+T_{15}+T_{18}-\nonumber\\
&&-\left(I_{6}-I_{12}\right)\frac{T_{27}}{2\rho_{2}}-\left[\frac{I_{2}+I_{3}}{2}+\left(\frac{a}{2}-b\right)\frac{aI_{8}}{\rho_{2}^2}\right]\frac{T_{32}}{2\rho_{2}}+\frac{I_{6}T_{33}}{2\rho_{2}},\ \ \ S_{25}=2S_{15}+\frac{I_{12}S_{27}}{2\rho_{2}}-\left(a-b\right)\frac{T_{22}}{2\rho_{2}}-\nonumber\\
&&-\left[R_{12}+\left(a-b\right)\frac{3I_{8}}{\rho_{2}}\right]\frac{T_{27}}{2\rho_{2}}-\left[\left(3a-\frac{13b}{2}\right)I_{6}-\frac{bI_{12}}{2}\right]\frac{T_{32}}{2\rho_{2}^2}+\left[\frac{R_{12}}{2}+\left(3a-\frac{19b}{4}\right)\frac{I_{8}}{\rho_{2}}\right]\frac{T_{33}}{\rho_{2}},\nonumber\\
&&T_{25}=-\left(a-\frac{b}{2}\right)\frac{S_{22}}{\rho_{2}}+\left(R_{12}-\frac{bI_{8}}{\rho_{2}}\right)\frac{S_{27}}{2\rho_{2}}+\frac{\rho_{1}T_{10}}{\rho_{2}}+T_{15}+T_{20}+\frac{I_{12}T_{27}}{2\rho_{2}}+\left[\frac{3bR_{12}}{4\rho_{2}}+\frac{\rho_{1}I_{1}}{\rho_{2}}-\frac{I_{2}-I_{3}-I_{5}}{2}\right]\frac{T_{32}}{\rho_{2}}+\nonumber\\
&&+\left(a^2-b^2\right)\frac{I_{8}T_{32}}{2\rho_{2}^3}-\frac{I_{12}T_{33}}{2\rho_{2}},\ \ \ S_{26}=-\frac{I_{8}S_{27}}{\rho_{2}}-\left(R_{12}+\frac{aI_{8}}{\rho_{2}}\right)\frac{T_{32}}{2\rho_{2}},\ \ \ T_{26}=T_{22}-\frac{I_{6}T_{32}-I_{8}T_{33}}{\rho_{2}},\nonumber\\
&&S_{28}=S_{27}-\left(\frac{a}{2}-b\right)\frac{T_{32}}{\rho_{2}},\ \ \ T_{28}=T_{27}-T_{33},\ \ \ S_{29}=S_{27}-\left(\frac{a}{2}-b\right)\frac{T_{32}}{\rho_{2}},\ \ \ T_{29}=T_{27}-T_{33},\nonumber\\
&&S_{30}=0,\ \ \ T_{30}=T_{32},\ \ \ S_{31}=0,\ \ \ T_{31}=T_{32},\ \ \ S_{32}=0,\ \ \ S_{33}=-\frac{aT_{32}}{2\rho_{2}},\ \ \ S_{34}=S_{27}+\frac{bT_{32}}{2\rho_{2}},\ \ \ T_{34}=0,\nonumber\\
&&4a\left[2\rho_{2}\left(\rho_{1}S_{10}-\rho_{2}S_{15}+\rho_{2}S_{18}\right)+\rho_{2}\left(I_{6}S_{27}+bT_{22}\right)+\left(a-b\right)I_{8}T_{27}\right]+a\left[2\rho_{2}R_{2}+\left(a+3b\right)I_{6}+aI_{12}\right]T_{32}+\nonumber\\
&&+2\left[2\rho_{2}\left(a-b\right)R_{12}+2\rho_{2}^2\left(I_{2}-I_{3}-2I_{5}\right)+a\left(b-2a\right)I_{8}\right]T_{33}=0,\ \ \ I_{6}T_{32}-I_{8}T_{33}=0.\nonumber
\eea}


\end{document}